\documentclass[final,5p,times,twocolumn]{elsarticle}

\usepackage{amssymb}
\usepackage{amsmath}

\usepackage{lineno}
\usepackage{graphicx}
\usepackage{hyperref}
\hypersetup{
    colorlinks=true,
    linkcolor=blue,
    filecolor=magenta,      
    urlcolor=cyan,
    }

\usepackage[dvipsnames]{xcolor}
\usepackage{color}
\definecolor{junglegreen}{rgb}{0.16, 0.67, 0.53}

\usepackage[capitalise]{cleveref}

\newcommand{\R}{\mathbb{R}}

\newcommand{\grad}[0]{\nabla}

\newcommand{\abs}[1]{\lvert {#1} \rvert}

\newcommand{\BigO}[1]{\mathcal{O}(#1)}

\newcommand{\transpose}[1]{{#1}^T}
\newcommand{\inv}[1]{{#1}^{-1}}

\newcommand{\obj}[0]{\mathcal{L}}
\newcommand{\adjoint}[1]{\overline{#1}}
\newcommand{\kloc}[0]{k^e}
\newcommand{\kglob}[0]{K^e}
\newcommand{\dofel}[0]{\textrm{dof}^e}

\newcommand{\gfga}{\texttt{GF-GA}}
\newcommand{\gfcobyla}{\texttt{GF-COBYLA}}
\newcommand{\fdmma}{\texttt{FD-MMA}}
\newcommand{\namma}{\texttt{NA-MMA}}
\newcommand{\gbmma}{\texttt{GB-MMA}}

\journal{Automation in Construction}

\begin{document}

\begin{frontmatter}

\title{A differentiable structural analysis framework for high-performance design optimization}

\author[mit-affiliation]{Keith J. Lee\corref{corresponding-author}}
\cortext[corresponding-author]{Corresponding author}
\ead{keithjl@mit.edu}
\author[ethz-affiliation]{Yijiang Huang}
\author[mit-affiliation]{Caitlin T. Mueller}

\address[mit-affiliation]{Massachusetts Institute of Technology, Cambridge, USA}
\address[ethz-affiliation]{Swiss Federal Institute of Technology (ETH), Zürich, Switzerland}
  
\begin{abstract}
Fast, gradient-based structural optimization has long been limited to a highly restricted subset of problems---namely, density-based compliance minimization---for which gradients can be analytically derived. 
For other objective functions, constraints, and design parameterizations, computing gradients has remained inaccessible, requiring the use of derivative-free algorithms that scale poorly with problem size.
This has restricted the applicability of optimization to abstracted and academic problems, and has limited the uptake of these potentially impactful methods in practice.
In this paper, we bridge the gap between computational efficiency and the freedom of problem formulation through a differentiable analysis framework designed for general structural optimization.
We achieve this through leveraging Automatic Differentiation (AD) to manage the complex computational graph of structural analysis programs, and implementing specific derivation rules for performance critical functions along this graph.
This paper provides a complete overview of gradient computation for arbitrary structural design objectives, identifies the barriers to their practical use, and derives key intermediate derivative operations that resolves these bottlenecks.
Our framework is then tested against a series of structural design problems of increasing complexity: two highly constrained minimum volume problem, a multi-stage shape and section design problem, and an embodied carbon minimization problem.
We benchmark our framework against other common optimization approaches, and show that our method outperforms others in terms of speed, stability, and solution quality.
\end{abstract}

\begin{keyword}
  Automatic Differentiation \sep
  Structural Analysis \sep
  Structural Optimization \sep
  Computational Design
\end{keyword}

\end{frontmatter}


\section{Introduction}\label{sec:introduction}
Optimization can play a critical role in the design of high performance structures.
By defining a measurable objective to minimize, it enables a systematic navigation through the design space towards a low cost solution.
In practice, however, its use has long been limited to a specific subset of problem formulations.
At the core of this limitation is the ability to quickly compute the gradient (sensitivities) of the objective with respect to the design variables.

Gradients enable exact knowledge of the underlying objective landscape, allowing for a rapid descent towards an optimal solution with minimal function evaluations.
Although structural analysis is composed of common algebraic operations with known derivatives, the complexity and scale of a typical design optimization problem makes their manual implementation unrealistic.
Even if the often thousands of individual partial derivatives of a structural analysis routine are manually implemented without error, the computational expense of actually evaluating such derivatives may outweigh the benefit of gradient-based optimization.

A notable exception is the field of topology optimization \cite{bendsoe1995Optimization,bendsoe2003Topology}, which is supported by an easy to compute, closed form expression of the gradient, enabling fast optimization at large problem scales.
However, this expression is restricted to a specific problem formulation: a minimum compliance (elastic strain energy) problem with element area variables and a total volume constraint.
Any deviation from this formulation loses the applicability of the known gradient expression, and reintroduces the barrier to fast optimization.
In these scenarios, one must resort to either finite-difference approximations of the gradient or the use of derivative-free optimization algorithms.

Neither approach is ideal, as they require significantly more function evaluations at each iteration, scaling poorly with both structure size and the number of design variables.
As a result, fast, scalable optimization has remained out of reach for many real-world problem formulations in structural design.
These include, among others: direct minimization of structural volume, stress and displacement constraints, and combining both geometric and cross section variables in the same problem.
Finding optimal solutions to these realistic problem formulations is increasingly important in light of the climate impact of the built environment \cite{worldgreenbuildingcouncil2019Bringinga}.

In this paper, we bridge the gap between gradient accessibility and flexible problem formulation to enable fast, scalable optimization for general structural design problems.
We achieve this through two components.
First, we leverage reverse-mode automatic differentiation (AD) to manage the directed graph of function calls in a structural analysis program, automating the laborious and error-prone process of tracking sensitivities between output and input values.
Second, we pair the mechanics of reverse-mode AD (backpropagation) with analytically-derived matrix derivatives of performance-critical functions to define specific differentiation rules;
these rules enable time and memory efficient gradient computations along the generated function graph.

After reviewing the state of the art in computational structural optimization, we provide a theoretical overview of gradient computation in matrix structural analysis, and the technical challenges in their practical use.
We then show how AD can be applied to overcome some of these challenges, and the derivation of our domain-specific differentiation rules that eliminate the remaining bottlenecks to fast gradient computation.
Our framework is tested on four design problems that vary in scale, the objective, and design parameterization.
We show how our framework provides a new baseline for high-performance optimization, enabling direct solutions to tangible design objectives.

\section{Related work}\label{litreview:related-work}
\subsection{(Computational) structural optimization}
The formal investigation of optimal structural systems has continued now for over a century, beginning with \citet{michell1904Limits}, who determined the ideal bar layout of truss structures through Maxwell's load path theorem \cite{maxwell1864Reciprocal}, from which new insights have occurred to the modern day \cite{mazurek2011Geometrical,baker2013Maxwell}.
Here, ``optimal'' is considered as the minimum theoretical volume required to safely construct the final design.
Although originally motivated by material economics, the weight of structural systems is now equally important when considering the environmental cost of building materials \cite{worldgreenbuildingcouncil2019Bringinga,fang2023Reducing}.
This enduring motivation has resulted in numerous analytic expressions and optimality criteria for a range of structural design problems, including beams \cite{barnett1961MinimumWeight,shield1970Optimal} and entire floor grillage layouts \cite{prager1977Optimala,prager1970Optimization,kirsch1986Optimal,bolbotowski2018Design,whiteley2023Engineering}.
An extensive overview of different optimization formulations is provided by \citet{rozvany1995Layout} and more recently by \citet{mei2021Structural}.

Advancements in computational power and accessibility enabled a broader range of problems to be addressed through numeric optimization, rather than analytic derivations of case-specific optimality criteria, including low-carbon beam design \cite{ismail2021Shaped,mayencourt2019Structural}, global geometry of frames and trusses \cite{preisinger2011Evolutionary,vierlinger2013Framework}, and lightweight surface structures \cite{bletzinger1997Generalized,bletzinger2001Structural}.
Although this new accessibility has been an inarguable benefit to the field, the difficulties in computing gradients for arbitrary structural design problems have left the actual process of optimization to derivative-free algorithms.
Popular approaches include direct-search methods that approximate the local design space \cite{powell1994Direct,powell2009BOBYQA}, or a broad selection of evolutionary or meta-heuristic algorithms that rely on stochastic sampling of parameters \cite{kaveh2017Advances,kicinger2005Evolutionary,xie1993Simple,munk2015Topology}.
Both approaches ultimately discard the gradient (that exists but is difficult to compute) in favor of repeated structural analysis and evaluation of the objective.
Not only does this scale poorly with problem size, in the case of evolutionary algorithms, their stochastic nature typically results in vastly different solution geometries which may bear little resemblance to the initial design.
Although this can enable diverse and unexpected final solutions, it is ultimately difficult to control directly.
As noted in the introduction, the field of topology optimization suffers from the opposite problem, in which gradients are accessible at the cost of a restrictive problem formulation.
In this paper, we show how we eliminate this longstanding compromise, primarily through the use of Automatic Differentiation.

\subsection{Automatic Differentiation (AD) and structural design}\label{litreview:ad}
Automatic---or Algorithmic---Differentiation is a method of computing exact derivatives of arbitrary mathematical functions expressed as computer programs \cite{griewank1991Automatic,griewank2008Evaluating}.
By decomposing such programs into a directed graph of elementary operations with known derivatives, the sensitivity of the overall function can be computed through repeated applications of the chain rule.
Most commonly associated with the training of machine learning models \cite{werbos1990Backpropagation}, it has found widespread adoption wherever the derivatives of complex programs are required \cite{margossian2019Review,hu2020DiffTaichi}.

In structural design, AD has enabled fast optimization for equilibrium-based form-finding frameworks, such as the Force Density Method (FDM) \cite{schek1974Force}.
\citet{cuvilliers2016Gradientbased} developed a framework for the inverse form-finding of funicular structures that best approximate a freeform target geometry, which has further been extended to support arbitrary design objectives and constraints \cite{pastrana2024Arpastrana,burke2023FDMremote}.
For mixed compression-tension structures, solutions to Combinatorial Equilibrium Modeling (CEM) problems have been accelerated by the use of AD \cite{oleohlbrock2016Combinatorial,pastrana2022Constrained}.
However, both FDM and CEM are specialized frameworks that solve for equilibrium geometries of a restricted subset of topologies---they cannot be used to \emph{analyze} an arbitrary structure at a given state during optimization.
Recently, \citet{wu2023Framework} incorporated AD, GPU programming, and code vectorization in compliance-based shape optimization using the Direct Stiffness Method \cite{kassimali2012Matrix}, enabling fast gradient computation in a general structural analysis framework.

\subsection{Research gap and opportunities}
In this paper, we present an analysis framework that dissolves the long-standing compromises that have limited the applicability of high-performance structural optimization.
Direct-search and evolutionary optimizers allow for a wide range of problem formulations at the cost of speed or design control; we maintain this flexibility but provide a fast and systematic navigation through the design space through gradient-based optimization.
Topology optimization sets the standard for fast and scalable structural optimization at the cost of a rigid problem formulation; we remove this limitation while maintaining the ease of gradient computation.
The efficiency of AD in structural design has already been proven for equilibrium modeling techniques such as FDM and CEM; we extend this utility to a broader class of structural systems by using a general analysis framework.

Our work shares the most similarity with \citet{wu2023Framework} due to the combination of AD and a general analysis framework.
However, this generalization was not extended to a freedom in problem formulation, relying instead on a compliance-based objective for fast gradient computation;
their work can be considered as an extension of topology optimization for spatial design variables.
In this paper, we enable generality in all aspects of design optimization: the type of structure, design variables, objectives, and constraints that can be differentiated.

\section{Methodology} \label{sec:methodology}
This paper focuses on bar-abstracted structures that are analyzed using the Direct Stiffness Method, which include trusses, beams, and frames (subject to axial, shear, and bending forces).
We begin with an overview of this method in \cref{sec:structural-analysis}, derive the analytic derivatives for key subprocesses, and identify the technical barriers of computing such derivatives in practice.
In \cref{sec:ad-and-adjoints}, we provide an overview of Automatic Differentiation and backpropagation in relation structural analysis, and derive the critical intermediate sensitivities (adjoints) that overcome the identified barriers.

\subsection{Structural analysis and optimization} \label{sec:structural-analysis}
Optimizing any objective that is dependent on structural behavior requires structural analysis to be performed at some point in the function evaluation.
In this paper, we consider linear elastic Finite Element Analysis (FEA) using the Direct Stiffness Method for bar structures.
Tracing its roots to the 1950s \cite{livesley1956Automatic,turner1956Stiffness}, the Direct Stiffness Method has seen lasting utility in the field of structural design for its relative ease of implementation and speed.
We provide a summary of key analysis steps in this section, and refer unfamiliar readers to \citet{kassimali2012Matrix} for complete theoretical details;
for those already familiar, the following section acts to define our chosen notation for important intermediate variables.

\subsubsection{Structural analysis overview} \label{sec:structural-analysis-overview}
A structural model is defined by edges with section properties (material stiffness, area, bending/torsional stiffness) connected at nodes with independent translational and rotational degrees of freedom.
Given an external set of loads, \(p\), applied at these nodes and fixed degrees of freedom that counteract them, we seek to solve the corresponding displacement vector of all degrees of freedom, \(u\), that satisfy the relation:
\begin{align}
  Ku &= p\label{eq:kup}\\
  u &= \inv{K}p \nonumber
\end{align}
Here, \(K\) is the global stiffness matrix that relates the displacement of a nodal degree of freedom, \(i\), to the force experienced at another, \(j\), through the topology and stiffness of the underlying graph of elements that connect all nodes together.
Naturally, if \(u \in \R^n\), then \(K \in \R^{n\times n}\).
The global stiffness matrix is a summation of \emph{elemental} stiffness matrices, \(\kglob\), that only relate the displacement-force relationship between the degrees of freedom corresponding to its two end nodes.
This matrix can be decomposed into two parts: a stiffness matrix in the local coordinate system of an element, \(\kloc\), and a coordinate transformation matrix, \(\Gamma\):
\begin{equation}
  \kglob = f(\Gamma, \kloc) = \transpose{\Gamma} \kloc \Gamma \label{eq:k-transformation}
\end{equation}

The local element stiffness matrix is a function of the element length and section properties.
For example, for a truss structure that resolves all displacements through axial deformation of elements:
\begin{equation}
  \kloc = f(E, A, L) = \frac{EA}{L} \begin{bmatrix}
    1 & -1 \\ -1 & 1
  \end{bmatrix} \label{eq:kloc}
\end{equation}
where \(E\) is the Young's modulus (material stiffness), \(A\) is the cross sectional area, and \(L\) is the element length.

In this local coordinate system, a single translational degree of freedom along the axis of the element exists at each end node, giving \(\kloc \in \R^{2\times 2}\).
The element transformation matrix, \(\Gamma\), translates this reference frame into a globally common (i.e., Cartesian) coordinate system through \cref{eq:k-transformation}.
For a truss element in a three dimensional structure:
\begin{equation}
  \Gamma = f(c_x, c_y, c_z) = \begin{bmatrix}
    c_x & c_y & c_z & 0 & 0 & 0\\
    0 & 0 & 0 & c_x & c_y & c_z
  \end{bmatrix} \label{eq:gamma}
\end{equation}
where \([c_x, c_y, c_z]\) are the components of the normalized element vector.
By computing \(K\) through the summation of elemental \(\kglob\)s at corresponding degree of freedom indices, we can solve for the global displacement vector \(u\) in \cref{eq:kup}, from which all other structural properties can be computed.
For example, we can determine the internal forces of a given element in its local reference frame by:
\begin{equation}
  F = f(\Gamma, \kglob, u) = \Gamma \kglob u_{\dofel} \label{eq:element-forces}
\end{equation}
where \(u_{\dofel}\) are the values in \(u\) corresponding to the element's degrees of freedom, \(\dofel\).

\subsubsection{Computing sensitivities}\label{sec:structural-sensitivities}
Consider an arbitrary objective function, \(\obj\), that depends on the structure's behavior, and thus the value of \(u\).
If our design variables, \(x\), affect the state of our structure (such as the position of a node or the cross sectional area of an element), then \(u\) is itself a function of \(x\), giving:
\[
  \obj = f(u(x), \dots)
\]

To minimize our objective, we should ideally compute its gradient, \(\grad_x \obj\), to find the direction in the design space that most efficiently improves the objective value.
This gradient can be readily computed through the recursive use of the chain rule.
Suppose \(x_i = A_j\), the cross sectional area of element \(j\) in a structure.
Then:
\begin{equation}
  \frac{d \obj}{d A_j} = \frac{d \kloc_j}{d A_j} \cdot \frac{d \kglob_j}{d \kloc_j} \cdot \frac{d K}{d \kglob_j} \cdot \frac{du}{dK} \cdot \frac{d\obj}{du} \label{eq:dobj-dA}
\end{equation}

\begin{figure*}
  \centering 
  \includegraphics[width = \textwidth]{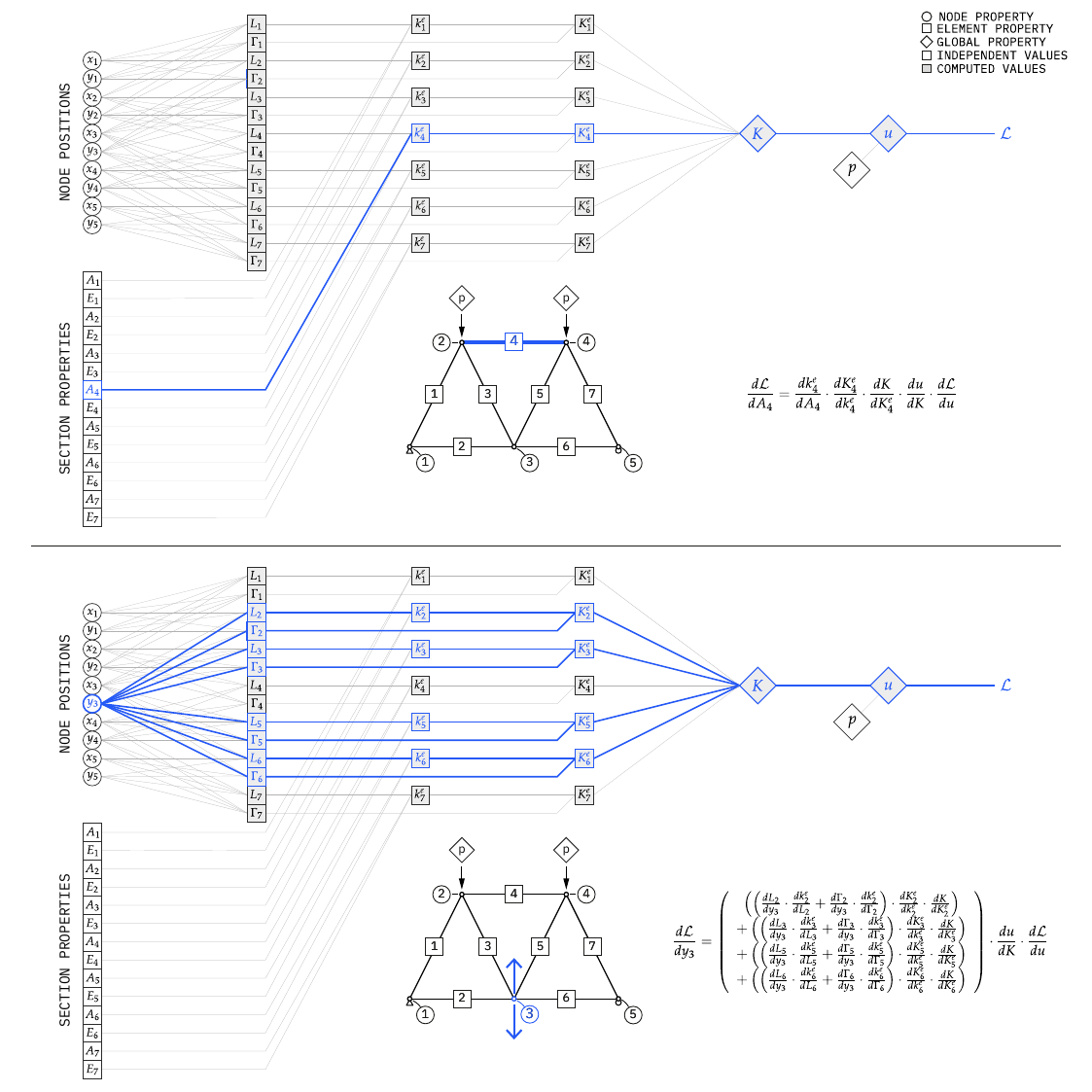}
  \caption{\label{fig:dag-optimization} The computation graph of a structural optimization problem for a seven-bar truss structure with an objective function, \(\obj(u,\dots)\). \emph{Top:} The effect of a single element parameter (cross sectional area) follows a single path along the graph, resulting in a relatively straightforward derivative calculation. \emph{Bottom:} A single spatial node parameter, however, affects multiple streams of intermediate computations; its derivative calculation reflects the sum of all these paths.}
\end{figure*}

Visualized in \cref{fig:dag-optimization}, the value of \cref{eq:dobj-dA} can be explicitly computed through standard calculus operations.
Why then, are such derivatives rarely, if ever, used in structural optimization?
The remainder of this section identifies and describes two key barriers to the effective computation and implementation of gradient-based structural optimization.

\subsubsection{Topology of operations}\label{sec:topology-operations}
In the previous section, we showed that an infinitesimal change in the cross sectional area of an element traces a single path through the intermediate computations that result in the global displacement vector and objective value.
Suppose now the decision variable is \(y_j\), or the vertical position of node \(j\) in a two-dimensional structure. 
It is clear that the position of a node affects more than the local coordinate stiffness matrix of a single element.
Rather, by changing the length and directional vector of \emph{all} elements that are connected to the given node, changes in \(y_j\) affect the value of \(\Gamma\) and \(\kloc\) for multiple parallel computational traces, as shown in the bottom half of \cref{fig:dag-optimization}.
Computing \(\frac{d\obj}{dy_j}\) amounts to computing each trace independently and summing their resulting values.

A structural optimization problem may have hundreds or thousands of nodes.
Although it is possible to define, compute, and sum these traces for each node, it is unrealistic to manually perform and is prone to error.
Further, the topological relations between nodes is dependent on an arbitrary numbering.
Should the numbering of nodes change, or if an element is added or removed in the design, a manual implementation of a spatial derivative may require complete rewriting.
Ideally, these traces are generated automatically, eliminating the chance of errors and adapting to changes in topology.

\subsubsection{Intermediate sensitivities}\label{sec:intermediate-sensitivities}
Suppose such function traces \emph{can} be generated automatically without error.
Is fast, gradient-based optimization now within reach? 
Not quite. 
One critical barrier is the following intermediate derivative:
\begin{equation}
  \frac{du}{dK} = \frac{d}{dK} \left(\inv{K}p\right) \label{eq:dudk}
\end{equation}
This value represents the sensitivity of each component of \(u\) with respect to each component of \(K\), amounting to a third order tensor:
\[\frac{du}{dK} \in \R^{n \times n \times n}\]
Symbolically, this can be solved through tensor calculus \cite{laue2020Simple}:
\begin{align*}
  \frac{du}{dK} &=- \transpose{(\inv{K}p)} \otimes \inv{K}\\
  &= -\transpose{u} \otimes \inv{K}
\end{align*}
where the Kronecker product, \(\otimes\), is used to represent the third-order tensor form of \cref{eq:dudk}, which can be explicitly written as:
\begin{equation}
  \frac{du}{dK} = -\begin{bmatrix}
  u_1\inv{K} & \dots & u_i\inv{K} & \dots & u_n\inv{K}
\end{bmatrix} \label{eq:dudk-expanded}
\end{equation}

When applied, however, \cref{eq:dudk} amounts to solving for the explicit inverse of \(K\), an operation with computational complexity \(\BigO{n^3}\).
Worse, \(K\) is typically a large sparse matrix, whereas \(\inv{K}\) is often dense, requiring significant memory to compute and store.
For structures of even moderate size, this explicit inversion creates either a time or memory bottleneck, leaving gradients accessible but impractical.

Another challenge is in computing the derivative of the global stiffness matrix, \(K\), with respect to an element's stiffness matrix:
\[
  \frac{dK}{d\kglob}
\]
Although seemingly innocuous compared to \cref{eq:dudk}, there is no compact mathematical representation for the assembly of the global stiffness matrix, resulting in challenges when expressing its derivative.
For example, most structural analysis programs assemble \(K\) by first initializing an empty sparse matrix, then iterating over each \(\kglob\):
\[
  K_{i,j} \leftarrow K_{i,j} + \kglob_{l,m} \quad \text{if} \quad \dofel_l = i, \dofel_m = j \label{eq:assemble-K}
\]
where \(\dofel\) are the indices of the degrees of freedom associated with the given element.
Although easily written as a computer program, it cannot be expressed through a closed form mathematical expression.
To solve this, one could store each \(\kglob\) as a hyper sparse matrix, \(\tilde{\kglob} \in \R^{n\times n}\), with nonzero entries corresponding to the element's degrees of freedom, giving:
\begin{equation}
  K = \sum_i \tilde{\kglob_i} \label{eq:assemble-K2}
\end{equation}
However, this approach again suffers from unnecessary memory consumption, and creates additional dimensionality challenges when working with the transformation matrix, \(\Gamma\).

The practical computation of gradients for structural optimization requires two components: 
(1) the automatic generation of the trace of functions \emph{from} the objective \emph{to} each design variable \(x_i\), and 
(2) time and memory efficient methods of computing intermediate sensitivities for complex functions.

\subsection{Automatic Differentiation and adjoint sensitivities}\label{sec:ad-and-adjoints}
We resolve the two critical barriers to gradient-based structural optimization through a combination of Automatic Differentiation (AD) and defining custom differentiation rules (adjoints) for bottleneck functions.
In this section, we provide a brief theoretical overview of AD and then provide the derivations of our specific adjoints.

\subsubsection{Automatic Differentiation}\label{sec:ad}
It is not a coincidence that our function trace diagram in \cref{fig:dag-optimization} resembles that of abstract drawings of neural networks in machine learning (ML).
Both our framework and the training of ML models requires a gradient computation through a densely connected graph of functions from input parameters to output objective values.
As such, we leverage the same underlying principle of ML model training for structural optimization: reverse-mode AD and backpropagation.

In short, AD takes in an arbitrary computer program that expresses a mathematical process, and generates the directed graph of function traces between inputs, intermediate values, and outputs.
Computing the gradient begins with the forward pass, evaluating and storing each intermediate function value between the input, \(x\), and output, \(f(x)\).
To compute the gradient, this graph is traversed backwards \emph{from} the output through all connected pathways \emph{to} the input parameters, with a repeated application of the chain rule to accumulate the overall sensitivity of the function.
For example, consider the nested function:
\begin{equation}
  y(x) = f(g(x), h(x))
\end{equation}
Computing \(y(x)\) follows two function traces: \(x \rightarrow g \rightarrow f\) and \(x \rightarrow h \rightarrow f\). Correspondingly, its derivative is the sum of applying the chain rule to each trace:
\begin{equation}
  \frac{dy}{dx} = \frac{dg}{dx}\cdot\frac{df}{dg} + \frac{dh}{dx}\cdot\frac{df}{dh} \label{eq:dydx}
\end{equation}
In practice, the number of individual traces corresponding to a single input may number in the hundreds or thousands; the use of AD enables an automatic propagation and summation of intermediate gradients through these pathways.

The actual derivative values are also computed automatically. 
Provided that they are mathematical in nature, any function can be decomposed into a set of elementary operations with known derivatives: addition, multiplication, exponentiation, etc. 
It suffices then to repeatedly apply known derivation rules across the entire function trace until the desired gradient is reached.
The ``mathematical'' requirement of a program undergoing AD is surprisingly broad:
conditional statements such as if/else clauses are readily differentiated by simply considering the clause that is active in that instance; loops can be unrolled into a linear set of operations and also differentiated.
Generally, to enable AD, it suffices to write the program as a set of \emph{pure} functions, where: (1) functions take in numerical inputs and return numerical outputs, (2) functions do not modify existing variables, and (3) all operations are deterministic.
By writing our structural analysis program in a pure functional approach, we can leverage AD to solve our first problem outlined in \cref{sec:topology-operations}. 

\subsubsection{Adjoint sensitivity} \label{sec:adjoints}
An important distinction of reverse-mode AD is that it does not compute intermediate derivative values individually and then perform a final multiplication and summation, as is suggested by \cref{eq:dydx}.
Rather, at each depth in the function graph between an output and input, it accumulates the sensitivity \emph{up to} the intermediate function's outputs in a process known as backpropagation.
For example, if a function \(\obj (x)\) relies on some intermediate value computed by a function \(g\), we denote with a bar, \(\adjoint{g}\), its adjoint variable, or the gradient of the objective with respect to each output value of \(g\):
\[\adjoint{g} = \frac{d\obj}{dg}\]

If the inputs to \(g\) are not the independent variables \(x\), but rather the output of yet another intermediate function \(h\), we continue the process of backpropagation by applying the chain rule with the previously computed adjoint:
\[\adjoint{h} = \frac{d\obj}{dh} = \frac{dg}{dh}\adjoint{g}\]
This process is repeated, traversing backwards through the function graph, until reaching the adjoint of the independent variables, which is the gradient of the overall function:
\[\adjoint{x} = \grad_x \obj\]

Backpropagation is perhaps counterintuitive to the conventional understanding of differential calculus, where one typically considers the resulting change of an output given an infinitesimal change of an input value.
In fact, this more intuitive understanding of gradient computation is the underlying principle of forward-mode AD \cite{martins2021Engineering}, where differential changes of inputs are tracked along the computational graph towards all output values.
In reverse-mode AD, we rather perturb an \emph{output} value and find the equivalent input values that would cause such a perturbation through a backwards traversal of the graph.

This fundamental difference provides a general guideline on which AD method is most efficient for optimization: forward-mode AD is preferred when there are much fewer inputs than outputs, as the sensitivity of all outputs with respect to a single input can be computed in a single pass. 
On the other hand, since backpropagation begins at an output node, it is ideal in scenarios when the number of inputs exceed the number of outputs.
For structural optimization, reverse-mode AD is the natural candidate, as our design objectives are typically scalar (such as compliance or volume), whereas the number of inputs can grow rapidly with the size of the structure (the number of nodes and elements).

Revisiting \cref{eq:dobj-dA}, we can rewrite the derivative expression as a series of backpropagation steps:
\begin{subequations} \label{eq:dobj-dA-adjoint}
  \begin{align}
    \adjoint{\obj} &= 1\\
    \adjoint{u} &= \frac{d\obj}{du} \adjoint{\obj}\\
    \adjoint{K} &= \frac{du}{dK} \adjoint{u} \label{eq:problem-child}\\
    \adjoint{\kglob_j} &= \frac{dK}{d\kglob_j} \adjoint{K}\\
    \adjoint{\kloc_j} &= \frac{d\kglob_j}{d\kloc_j} \adjoint{\kglob_j}\\
    \frac{d\obj}{dA_j} = \adjoint{A_j} &= \frac{d\kloc_j}{dA_j} \adjoint{\kloc_j}
  \end{align}
\end{subequations}

Although the end result is unchanged, leveraging the form of \cref{eq:dobj-dA-adjoint} is critical for fast and memory efficient computation of gradients, especially when intermediate variables are high dimensional vectors and matrices.
A key driver of this efficiency is that the adjoints of many matrix-valued operations are typically functions of their original inputs---values that have already been computed and stored in the forward function evaluation.
As an example, given an intermediate function that performs a matrix multiplication:
\[f(A,B) = AB\]
From which \(A\) and \(B\) are computed intermediated values, their adjoint variables are:
\begin{subequations} \label{eq:AB}
  \begin{align}
    \adjoint{A} &= \frac{df}{dA} \adjoint{f} = \adjoint{f}\transpose{B}\\
    \adjoint{B} &= \frac{df}{dB} \adjoint{f} = \transpose{A} \adjoint{f}
  \end{align}
\end{subequations}

This can be verified by explicitly performing element-wise matrix multiplication and summing the partial sensitivities of \(f\) with respect to each component of \(A\) or \(B\);
readers can also refer to \citet{giles2008Extended} for a complete overview of matrix adjoints.
The primary benefit here is that \(A\) and \(B\) have already been computed in the forward evaluation of the objective function, and do not need to be reevaluated when computing the gradient.

AD packages in many programming languages already come with backpropagation rules for many common matrix operations; the remainder of this section will focus on leveraging our knowledge of the structural analysis pipeline to define our own adjoint sensitivities for critical operations to optimize the performance of gradient computation.

\paragraph{Nodal displacement}
We first return to the computational bottleneck defined in \cref{eq:dudk}:
\[
  \frac{du}{dK} = -\transpose{u} \otimes \inv{K}
\]
As mentioned, explicitly solving for the inverse of a large sparse matrix \(K\) is computationally expensive and memory intensive.
And unlike other common matrix adjoints, \(\inv{K}\) has not been computed in the forward pass of the function; instead, one would typically use an iterative solver or matrix factorization to directly solve for \(u\).
However, we know that during backpropagation, the actual value we seek is rather:
\begin{equation*}
  \adjoint{K} = \frac{du}{dK} \adjoint{u} = \left(-\transpose{u} \otimes \inv{K}\right) \adjoint{u}
\end{equation*}
which is a first mode tensor contraction \cite{cichocki2009Introduction}, equivalent to:
\[
  \adjoint{K} = -\begin{bmatrix}
    u_1\inv{K}\adjoint{u} & \dots & u_i\inv{K}\adjoint{u} & \dots & u_n\inv{K}\adjoint{u}
  \end{bmatrix}
\]
allowing for the compact form:
\begin{align}
  \adjoint{K} &= -\left(\inv{K}\adjoint{u}\right)\transpose{u} \nonumber\\
  &= -\left(\inv{K}\adjoint{u}\right)\transpose{\left(\inv{K}p\right)} \label{eq:deconstructed-kbar}
\end{align}

In this form, the adjoint of \(K\) is simply the outer product between the solutions to two linear systems, eliminating the need to compute an explicit matrix inverse and making accessible the wealth of linear solving strategies available in common scientific programming packages.

There are further benefits to this adjoint form.
First, \(\inv{K}p\) is merely the nodal displacement vector, \(u\), already computed in the forward function evaluation and can be reused during backpropagation.
Second, if we perform a factorization of \(K\) in the forward pass to solve for the nodal displacements, we can reuse this factorization in the backwards pass for \(\inv{K} \adjoint{u}\), allowing for a more efficient linear solve.
Third, since the sparsity pattern of \(K\) is dependent on structural topology and does not change for a given design, symbolic factorization strategies can also be implemented and reused between all passes for more efficient memory usage \cite{davis2004Algorithm}.

\paragraph{Global stiffness matrix assembly} \label{eq:K-assembly}
We take a step backwards in the function graph and consider our second bottleneck:
\begin{equation}
  \adjoint{\kglob} = \frac{dK}{d\kglob}\adjoint{K} \label{eq:kglob-pseudo-bar}
\end{equation}

Recall that the assembly of the global stiffness matrix cannot be expressed as a pure mathematical function, and thus its adjoint sensitivity requires some consideration.
We focus on the direct relationship between the inputs (the individual entries of an element's stiffness matrix, \(\kglob\)) with respect to the overall output (the global stiffness matrix, \(K\)).
Consider a single entry in an element stiffness matrix, \(\kglob_{l, m}\), corresponding to global degrees of freedom \(\dofel_l, \dofel_m\).
The derivative of the global stiffness matrix with respect to this single entry is then:
\begin{equation}
  \frac{dK_{i,j}}{d\kglob_{l,m}} = \begin{cases}
  1  \; \textrm{if} \; i = \dofel_i, j =  \dofel_m\\
  0 \; \textrm{otherwise} \label{eq:dK-dKelm}
\end{cases}
\end{equation}
The adjoint sensitivity of \(\kglob_{l,m}\) is then:
\begin{equation}
  \adjoint{\kglob_{l,m}} = \sum \frac{dK}{d\kglob_{l,m}} \odot \adjoint{K} \label{eq:Kelm-bar}
\end{equation}
where \(\odot\) is the Hadamard (element-wise) product.
But considering our knowledge of \cref{eq:dK-dKelm}, this is nothing more than the entry in \(\adjoint{K}\) at \(\dofel_l, \dofel_m\).
Repeating this process for each entry in \(\kglob\) gives the final adjoint sensitivity:
\begin{equation}
  \adjoint{\kglob} = \adjoint{K}_{\dofel, \dofel} \label{eq:kglob-bar}
\end{equation}

Here, \(\adjoint{K}_{\dofel, \dofel}\) denotes the entries of \(\adjoint{K}\) corresponding to the degrees of freedom of a given element.
Although the assembly of \(K\) cannot be expressed in a compact mathematical form, we leverage our understanding of the structure of \(K\) and \(\kglob\) to nonetheless define an efficient adjoint.

\paragraph{Coordinate transformation}
For completion, we define the closed form adjoint sensitivities for the transformation of the element stiffness matrix from local to global coordinates first defined in \cref{eq:k-transformation}:
\[
  \kglob = \transpose{\Gamma}\kloc\Gamma
\]

Although we can simply apply \cref{eq:AB} recursively, we leverage the fact that \cref{eq:k-transformation} is the quadratic form of \(\kloc\) and \(\Gamma\), giving the known adjoints \cite{giles2008Extended}:
\begin{align}
  \adjoint{\Gamma} &= \kloc \Gamma \transpose{\adjoint{\kglob}} + \transpose{\kloc} \Gamma \adjoint{\kglob} \label{eq:gamma-bar0}\\
  \adjoint{\kloc} &= \Gamma \adjoint{\kglob} \transpose{\Gamma} \label{eq:kloc-bar}
\end{align}
Since \(\kloc\) is symmetric, we can further reduce \cref{eq:gamma-bar0} to:
\[
  \kloc \Gamma \left(\transpose{\adjoint{\kglob}} + \adjoint{\kglob}\right)
\]

\subsection{Special case: the compliance adjoint}\label{sec:compliance-equivalent}
The term \emph{adjoint} in the context of structural optimization is most associated with the field of topology optimization, in which the objective is to find the optimal distribution of material density in a fixed set of elements that minimizes structural compliance:
\begin{align}
  \min_A \quad & \obj = C = u^Tp\\
  \text{s.t.} \quad & V \leq V_{\max} \nonumber
\end{align}
For this specific formulation, \citet{bendsoe1995Optimization} in the 1990s derived the explicit gradient for compliance with respect to an element's area (or density) using the adjoint method \cite{johnson2021Notes}, a form of problem augmentation akin to Lagrangian multipliers in constrained optimization.
The result is a closed form expression for the derivative:
\begin{equation}
  \frac{dC}{dA} = -\transpose{u_{\dofel}} \frac{d\kglob}{dA} u_{\dofel} \label{eq:dC-dA}
\end{equation}
Where \(\frac{d\kglob}{dA}\) is simply \(\kglob / A\).
Although \cref{eq:dC-dA} has long enabled the successful use of gradient based optimization in high performance structural design, it has also restricted the types of problems that one can consider.
Only recently has compliance-based optimization been extended to spatial variables using AD \cite{wu2023Framework}, but again, their formulation relies on a slightly generalized adjoint solution to the structural compliance objective:
\begin{equation}
  \frac{dC}{dx} = -\transpose{u_{\dofel}} \frac{d\kglob}{dx} u_{\dofel} \label{eq:dC-dx}
\end{equation}
where \(x\) can be generalized to other section properties or nodal positions.
Our framework makes no assumption on the objective to differentiate, while maintaining computational performance.
We demonstrate generality by deriving the derivative expression in \cref{eq:dC-dA} using backpropagation and our derived intermediate adjoints.
Given:
\[
  \obj = C = \transpose{u}p  
\]
then:
\[\adjoint{u} = \frac{dC}{du} = p\]
substituting into \cref{eq:deconstructed-kbar}:
\[\adjoint{K} = -\left(\inv{K}p\right)\transpose{\left(\inv{K}p\right)} = -u\transpose{u}\]
then for a specific element, we have:
\[\adjoint{\kglob} = \adjoint{K}_{\dofel, \dofel} = -u_{\dofel}\transpose{u_{\dofel}}\]
and since \(A\) does not affect the coordinate transformation matrix, we continue to \cref{eq:kloc-bar}:
\[\adjoint{\kloc} = -\Gamma \left(u_{\dofel} \transpose{u_{\dofel}}\right) \transpose{\Gamma}\]
leading to the last step of backpropagation:
\[\adjoint{A} = \frac{dC}{dA} = -\sum \frac{d\kloc}{dA} \odot \Gamma \left(u_{\dofel} \transpose{u_{\dofel}}\right) \transpose{\Gamma}\]
which is equivalent to \cref{eq:dC-dA}, with the difference that our approach first transforms global displacements back into the element's local coordinate system before multiplication and summation.

\subsection{Implementation} \label{sec:implementation}
In the previous section, we showed how a combination of AD and explicit adjoint derivations can enable efficient gradient-based optimization for arbitrary structural optimization problems.
Our methodology is implemented in a contained structural optimization package written in the Julia programming language with the following key dependencies:
\begin{enumerate}
  \item Asap.jl \cite{lee2024Asapa} for core structural analysis
  \item Zygote.jl \cite{innes2019Don} for Automatic Differentiation
  \item ChainRulesCore.jl \cite{white2024JuliaDiff} to define adjoint sensitivities
  \item Nonconvex.jl \cite{tarek2023Nonconvex} to perform optimization
\end{enumerate}

The entirety of our codebase, including the scripts used in \cref{sec:results}, is available in a public repository \cite{Lee_DiffAnalysis_AIC2024}.

\section{Results} \label{sec:results}
Our framework is tested against a series of design optimization problems with varying objectives, constraints, and parameterizations. 
In \cref{sec:warren} and \cref{sec:spaceframe}, we also compare the performance of our framework to four other optimization approaches.
Unless otherwise stated, all units are in \([\text{m}, \text{kN}]\), and all examples are performed on an M1 Pro Macbook using a single thread.
We provide the total time rather than the number of iterations as a measure of optimization performance to serve as a reference point for future extensions of this work.

\subsection{Constrained minimum volume optimization I}\label{sec:warren}
\begin{figure}
  \centering
  \includegraphics{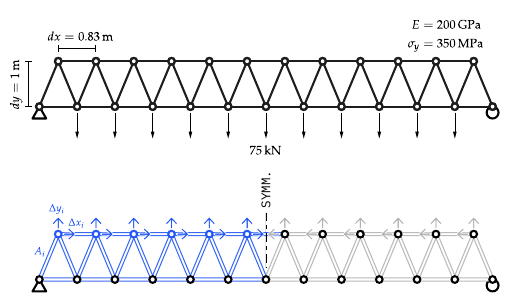}
  \caption{\label{fig:warren-prob-form} Problem formulation and variable definition for a minimum volume optimization problem.}
\end{figure}

We consider a simultaneous shape and area optimization of a planar truss to find the minimum volume solution that meets all stress and displacement constraints.
A ten meter span, simply supported Warren truss is split into twelve bays at an initial depth of one meter (\cref{fig:warren-prob-form}).
The problem is formalized as:
\begin{align}
  \min_{x} \quad & f(x) = V = \transpose{A}L \label{eq:min-volume}\\
  \text{s.t.} \quad & \abs{d_{yi}} \leq d_{\max} \quad \forall \; i \in 1,\dots, n_n \nonumber\\
  & \abs{\sigma_i} \leq \sigma_{\max} \quad \forall \; i \in 1,\dots,n_e \nonumber
\end{align}

Assuming that the nodes of the bottom chord are static to support a constant grade deck, our spatial design variables are assigned to the top nodes only, where for an element \(i\) we have two variables, \(\Delta x, \Delta y\), that dictate the change from the initial position.
We also include the cross section area of each element in our vector of variables.
Symmetry is enforced by assigning mirrored values to node positions across the vertical symmetry axis such that:
\begin{align*}
  \Delta x_L &= - \Delta x_R\\
  \Delta y_L &= \Delta y_R\\
  A_L &= A_R
\end{align*}
where \(L,R\) represent the left and right pairs of nodes and elements.
The only exception to this coupling is the top chord element that intersects the symmetry axis, which is assigned its own independent area variable without any coupling. Given six independent nodes with two degrees of freedom and 24 independent areas, we define our design vector:
\[x \in \R^{36} = \transpose{\begin{bmatrix} \Delta x_1 & \Delta y_1 & \dots & A_{24}\end{bmatrix}}\]

In the case of \cref{eq:min-volume}, the gradient of the objective function does not rely on structural analysis and is relatively straightforward to compute:
\begin{align*}
  \nabla_A V &= L\\
  \nabla_L V &= A
\end{align*}
Intuitively, we can minimize our objective by directly reduce the values in \(x\) corresponding to cross sectional areas and setting spatial values in \(x\) such that edge lengths are minimized.
The primary benefit of our method in this context is in computing the derivatives of the constraint functions, which do rely on a complete structural analysis on the state of the structure at each iteration.
The displacement constraint is directly computed from the solution to \cref{eq:kup}, from which gradients can be propagated using the adjoints defined in \cref{sec:adjoints}.
Computing the stress requires additional analysis. For a given element:
\begin{align}
  \sigma &= \frac{F}{A} \nonumber\\
  F &= \Gamma \kglob u_{\dofel} \label{eq:Fi}
\end{align}
Where \(F\) is a vector in \(\R^2\), from which the second component provides the proper magnitude and sign of axial force.
As all three terms that make up \cref{eq:Fi} are themselves functions of both spatial and cross section variables, we define the following adjoints for each variable:
\begin{align}
  \adjoint{\Gamma} &= \frac{dF}{d\Gamma}\adjoint{F} = \adjoint{F} \transpose{\left(\kglob u\right)} \label{eq:gamma-adjoint}\\
  \adjoint{\kglob} &= \left(\transpose{\Gamma}\adjoint{F}\right)\transpose{u} \label{eq:kglob-adjoint}\\
  \adjoint{u_{\dofel}} &= \transpose{\left(\Gamma \kglob\right)} \adjoint{F} \label{eq:u-adjoint}
\end{align}
Where the overall adjoint sensitivity of nodal displacements, \(\adjoint{u}\), is the resulting summation of element end node adjoint values shown in \cref{eq:u-adjoint}.
We refer the reader to \cite{giles2008Extended} for the general derivation of these adjoints. 
Given 25 nodes and 47 elements in the truss structure, a total of 72 constraints are specified, requiring at each optimization iteration the computation of a Jacobian matrix of size:
\[J_x \; c(x) \; \in \; \R^{36 \times 72}\]

The bounds of all spatial variables are set such that nodes do not cross each other in the \(x\) direction, and to ensure a non-zero truss depth in the \(y\) direction; cross sectional area variables are initialized to a large value to ensure a conservative starting position in the feasible region of the design space:
\begin{align*}
  -0.83 <\, \Delta x_0 &= 0 <0.83\\
  -1 <\, \Delta y_0 &= 0 \leq 1\\
  1\times 10^{-4} \leq\, A_0 &= 0.15 \leq 0.2
\end{align*}
For the constraints, we assume typical mild steel material properties and a vertical deflection limit of \(L/360\), giving:
\begin{align*}
  d_{\max} &= 2.78 \; \text{cm}\\
  \sigma_{\max} &= 350 \;\text{MPa}
\end{align*}

A total of five optimization algorithm-sensitivity combinations are applied to solve the same problem:
\begin{enumerate}
  \item \gfcobyla: A direct-search algorithm, Constrained Optimization BY Linear Approximation (COBYLA) \cite{powell1998Direct}
  \item \gfga: The Genetic Algorithm (GA) \cite{golberg1989genetic} with a population size of 100
  \item \fdmma: The Method of Moving Asymptotes (MMA) \cite{svanberg1987Method} algorithm using finite difference approximation of the gradients
  \item \namma: The MMA algorithm using generic AD without explicit adjoints to compute gradients
  \item \gbmma{} (our method): The MMA algorithm using AD and explicit adjoint functions to compute gradients
\end{enumerate}
The stopping criteria is set to a relative objective value tolerance of \(1 \times 10^{-6}\), or two minutes of computation.

\begin{figure}
  \centering
  \includegraphics{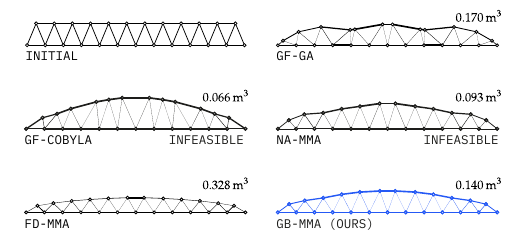}
  \caption{\label{fig:warren-solutions} Comparing the total optimization time with the final solution geometry and volumes. Red indicates a final solution that violates one or more constraints.}
\end{figure}

We show in \cref{tab:warren} and \cref{fig:warren-solutions} the overall results of the five optimization runs. 
From all five approaches, our method finds the lowest volume structure that remains in the feasible region of the design space. 
Although both \gfcobyla{} and \namma{} reach final states with a lower objective value, they occur in infeasible regions with nodal displacements greater than the allowable \(2.78\text{cm}\), and could not be resolved within the two minute stopping criteria.

\begin{table}
  \caption{\label{tab:warren} Warren truss optimization summary.}
  \resizebox{\columnwidth}{!}{%
  \begin{tabular}{lccccc}
  \hline
                       & GF-COBYLA & GF-GA  & FD-MMA & NA-MMA  & GB-MMA (ours) \\ \hline
  Time (s)             & 11.74      & 5.35   & 126.31 & 120.96 & 3.79   \\
  Volume ($\text{m}^3$) & 0.066    & 0.170 & 0.328  & 0.093   & 0.140 \\
  Feasible    & No        & Yes    & Yes    & No      & Yes    \\ 
  \hline
  \end{tabular}%
  }
\end{table}

Only the GA approach (\gfga) reaches a feasible solution close to our method within a similar timeframe, but with a distinctive final geometry unlike all other solutions.
This visual uniqueness is anticipated from a stochastic algorithm, which is not explicitly guided by the geometry of the underlying fitness landscape.
Although this independence from the fitness landscape may enable solutions with higher performance than the true local minimum, this cannot be guaranteed;
further, the non-determinism of GA and similar evolutionary algorithms makes replicability cumbersome.

\begin{figure}
  \centering
  \includegraphics{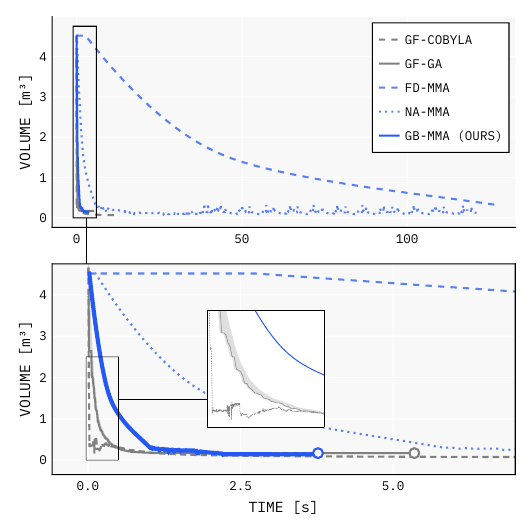}
  \caption{\label{fig:warren-traces} Optimization history of the minimum volume design problem. Our method converges to a feasible solution in the shortest time with the fewest iterations.}
\end{figure}

In \cref{fig:warren-traces}, we show the objective value history across all five optimization runs.
The upper plot shows how both \fdmma{} and \namma{} are effectively scaled versions of our method in the time axis, showing the significant speed benefits of explicit adjoint definitions for even relatively small optimization problems.
In the lower plot, we observe the smoothness of the objective value decrease between our method, \gfcobyla, and \gfga.
Although both gradient-free algorithms descend at a much rapid rate, they do so at the cost of constraint violation.
For \gfcobyla, after the initial near-vertical drop in the objective value, there is significant variation as the algorithm seeks to reestablish the performance landscape approximation and step towards feasible solutions.
Further, although the monotonically decreasing step function behaviour of \gfga{} suggests some level of smoothness, we show in gray bands the variation of the objective values for the entire population at each iteration; the dark line represents the best feasible solution within this population.

\subsection{Constrained minimum volume optimization II}\label{sec:spaceframe}
\begin{figure*}
  \centering
  \includegraphics[width=\textwidth]{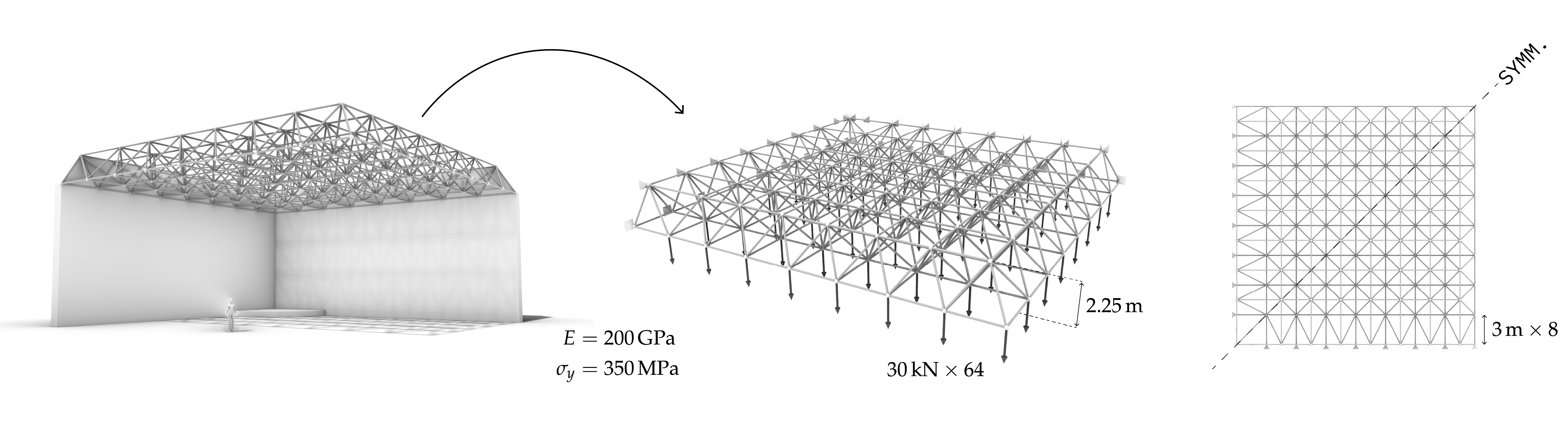}
  \caption{\label{fig:spaceframe-init} A cantilevered spatial truss roof structure consisting of \(n_n = 145\) nodes and \(n_e = 512\) steel elements.}
\end{figure*}
We consider the same minimum volume optimization problem as in \cref{sec:warren}, but at a larger scale.
A square aspect trussed roof structure supported on two edges is defined in \cref{fig:spaceframe-init}.
Covering a total area of \(24 \times 24=576\text{m}^2\), this structure is discretized into 145 nodes and 512 elements, with an overall depth of \(2.25\text{m}\).
Our design variables are the cross sectional areas of all elements as well as the vertical (\(z\)-axis) positions of the free nodes in the top chord.
As before, we enforce symmetry in the solution by coupling the vertical positions of nodes between pairs along a diagonal line of symmetry from the supported corner to the free corner;
the nodes along this symmetry line are not coupled.
We do not enforce symmetry in the areas of the elements to both test a larger problem scale and to evaluate the quality of solutions that result from our five optimization frameworks;
since the boundary conditions, loading, and geometry of our structure is symmetric, we should expect the distribution of element areas to also reflect this symmetry without explicitly specifying it.
The result is a design vector, \(x \in \R^{512}\), consisting of 28 spatial design variables and 484 area variables, with bounds and initial values:

\begin{align*}
  -1.25 \leq \Delta z_0 &= 0 \leq 4.5\\
  1 \times 10^{-3} \leq A_0 &= 0.1 \leq 0.2
\end{align*}

\begin{figure*}
  \centering 
  \includegraphics[width=\textwidth]{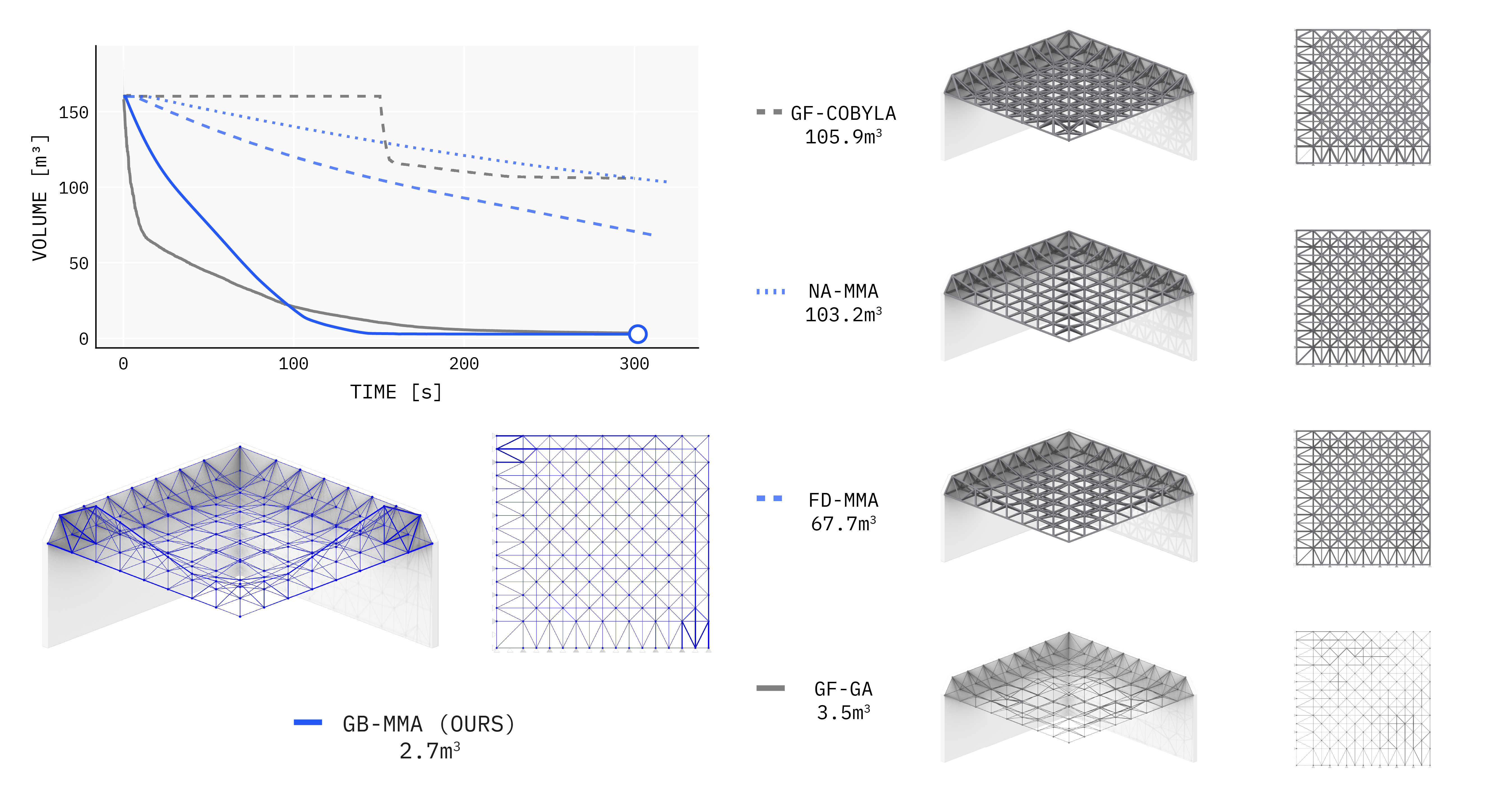}
  \caption{\label{fig:spaceframe-sols} Solutions to the spatial truss structure optimization problem. Our method finds the lowest volume solution by generating cantilevering support structures along the two free edges while minimizing the truss depth in the interior of the structure. Compared to \gfga, our method naturally distributes section areas evenly across the symmetry line.}
\end{figure*}

We take \(\sigma_{\max} = 350\text{MPa}\) and \(d_{\max}=8\text{cm}\) as the limits to our stress and displacement constraints.
These constraints are applied to all free nodes and elements, resulting in 596 total constraints in the optimization problem. A compute time limit of five minutes is set for all algorithms. The results for all five optimization processes are shown in \cref{fig:spaceframe-sols} and summarized in \cref{tab:spaceframe}.
For this problem, the five minute time limit was the stopping criterion for all solutions.

Observing the objective value history in \cref{fig:spaceframe-sols}, we see the effect of problem scale on the five optimization approaches tested in this section.
The Genetic Algorithm (\gfga) requires a single forward evaluation of both the objective and constraint functions at each iteration, scaled by population size (\(N=100\)), giving a total number of \(2N\) evaluations per iteration.
AD-based methods (our method and \namma) also perform a single forward call to both functions. 
However, as these methods require a gradient computation, the backpropagation of sensitivities from outputs to inputs must also be performed; this requires an additional computation for \emph{each} output of both functions, resulting in \(2 + 1 + n_c\) computations at each iteration, where \(n_c\) is the number of constraints. 
When performing finite differencing to compute gradients, as in \fdmma, this penalty increases further, as the relatively efficient output-wise backpropagation steps are replaced by more expensive forward computations for \emph{each} output with respect to \emph{each} input, leading to \(2 + n_x + n_xn_c\) forward evaluations at each iteration.

The most difficult to scale to complex optimization problems are gradient-free methods that seek to approximate the design space, such as \gfcobyla. 
In this case, a linear approximation of the objective landscape is generated by sampling the objective and each constraint at \(n_x + 1\) points around the current design variable value, \(x'\), and iterating upon this approximation.
At an upper bound, this requires \(2 + n_x + n_c(2+n_x)\) computations per iteration.

These challenges in computational efficiency are reflected in the trajectories of objective value histories in \cref{fig:spaceframe-sols}. 
All gradient-driven algorithms provide a monotonically decreasing objective function value, albeit at different rates depending on the manner of gradient extraction.
\gfga{}, which benefits from the fewest overall computations and a wider sampling of the design space, allows for the steepest initial decrease in the objective value, but suffers from slower convergence as it is unable to leverage the underlying design space geometry to directly find a local minimum.
Our method provides balance between computational efficiency and gradient-based navigation to converge to the overall optimal solution within the same timeframe.
The definition of custom adjoint functions is critical to our success. We note that in this example, the use of finite differencing (\fdmma) is actually \emph{more} efficient than using standard AD without adjoints (\namma), primarily due to the number of backpropagation steps required, resulting in repeated computations of \cref{eq:dudk}, the main performance bottleneck described in \cref{sec:adjoints}.

\begin{table}
  \caption{Spatial truss optimization summary.}
  \label{tab:spaceframe}
  \resizebox{\columnwidth}{!}{%
  \begin{tabular}{lccccc}
  \hline
                       & GF-COBYLA & GF-GA & FD-MMA & NA-MMA & GB-MMA (ours) \\ \hline
  Time (s)             & 300.2     & 300   & 313.9  & 321.4  & 302           \\
  Volume ($\text{m}^3$) & 105.9     & 3.51  & 67.6   & 103.2  & 2.73          \\
  Feasible             & No        & Yes   & Yes    & Yes    & Yes           \\ \hline
  \end{tabular}%
  }
\end{table}

Ultimately, only \gfga{} results in a solution that is within the same order of magnitude as our method.
However, despite the low final volume and satisfying all constraints, the resulting solution consists of a disorganized distribution of section areas and is structurally unintuitive.
Although our problem formulation does not enforce symmetry on the cross sectional area variables, our gradient-based approach naturally responds to the latent symmetric distribution of internal forces through the assigned section areas.
In contrast, since \gfga{} performs stochastic sampling of variables in the design space, cross sectional areas are also assigned randomly, resulting in a solution that does not reflect the symmetric nature of the structure and boundary conditions.
We argue that gradient-based optimization provides an implicit level of designer control, as it seeks a locally optimal solution near the initial conditions that the designer chooses.
Although this does not guarantee global optimality, it enables the degree in which an optimal solution deviates from the initial intended geometry, which may reflect implicit or difficult to quantify objectives, such as aesthetics.

In our method, we observe the emergence of a clear structural hierarchy: the collection of elements that make up the two free edges of the structure are shaped as cantilever beams, with the largest separation between the top and bottom chords occurring near the supports.
The interior region of the roof then forms a vault-like structure that is supported by either the walls or the stiffened free edges.
Here, we observe the interaction between efficient structural geometry (at the cost of longer elements) and the reduction of internal forces (enabling smaller areas).
In the case of \gfga{}, this structural hierarchy is not evident; rather, the solution is driven by \emph{any} combination of geometry and area that obtains a feasible design.

\subsubsection{Constructability-driven optimization}\label{sec:spaceframe-rationalization}
\begin{figure}
  \centering
  \includegraphics{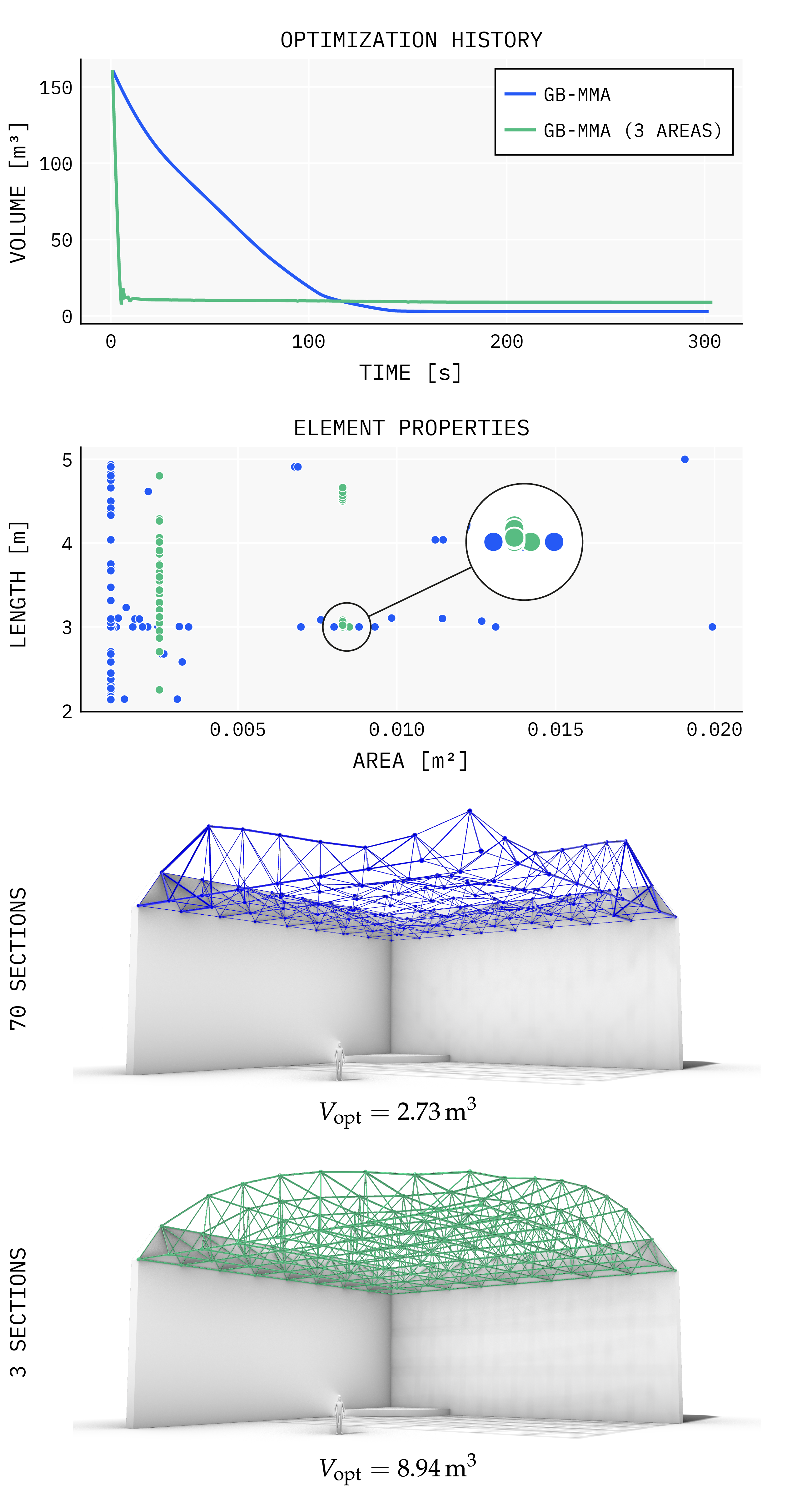}
  \caption{\label{fig:spaceframe-rationalization} Our method enables flexible parameterization of design variables to find locally optimal solutions while controlling the complexity of the final result. Here, we define only three cross section area variables that are applied to the top chord, web, and bottom chord elements respectively, and compare the result to the original solution using \gbmma. Note that the difference in area between the top and bottom chords is quite small at \(2.1\times 10^{-4}\,\text{m}^2\), and could also be grouped together with minimal effect on overall volume.}
\end{figure}

In the previous example, we allowed for independent sizing of each of the 484 elements to design a minimum volume structure.
Ultimately, our optimal solution (\gbmma) results in 70 unique cross sections, posing a significant logistical and cost challenge in fabrication and assembly.
Ideally, the number of unique sections---and the resulting complexity---should also be controllable by the designer.
Rather than rationalizing cross section areas as a post-processing step, we can leverage our framework to assign a single value to multiple structural variables to reduce the number of unique areas a priori, \emph{then} solving for a minimum volume solution.
For example, we can define a single area variable for three distinct element groups (top chord, web, and bottom chord), reducing the dimension of the design variable from 512 to 31.
Performing the same optimization problem defined in \cref{eq:min-volume}, we arrive at the second solution shown in \cref{fig:spaceframe-rationalization}.

Reflective of the much smaller design space, the three area formulation converges more rapidly, and results in an optimal geometry that is visually distinct from the previous solution, with a final volume that is approximately three times larger.
Whether this additional material cost justifies the reduction in construction complexity is dependent on the project and designer; our framework enables such comparisons to be made quickly and efficiently for effective decision making.
Further, the visual distinction between the initial and the three area solutions emphasizes that rationalization should \emph{not} be a post-processing strategy: the geometry that minimizes individually tailored element areas is not the same as one that minimizes a rationalized area formulation.

\subsection{Irregular frame design}\label{sec:funicularize}
\begin{figure}
  \centering
  \includegraphics{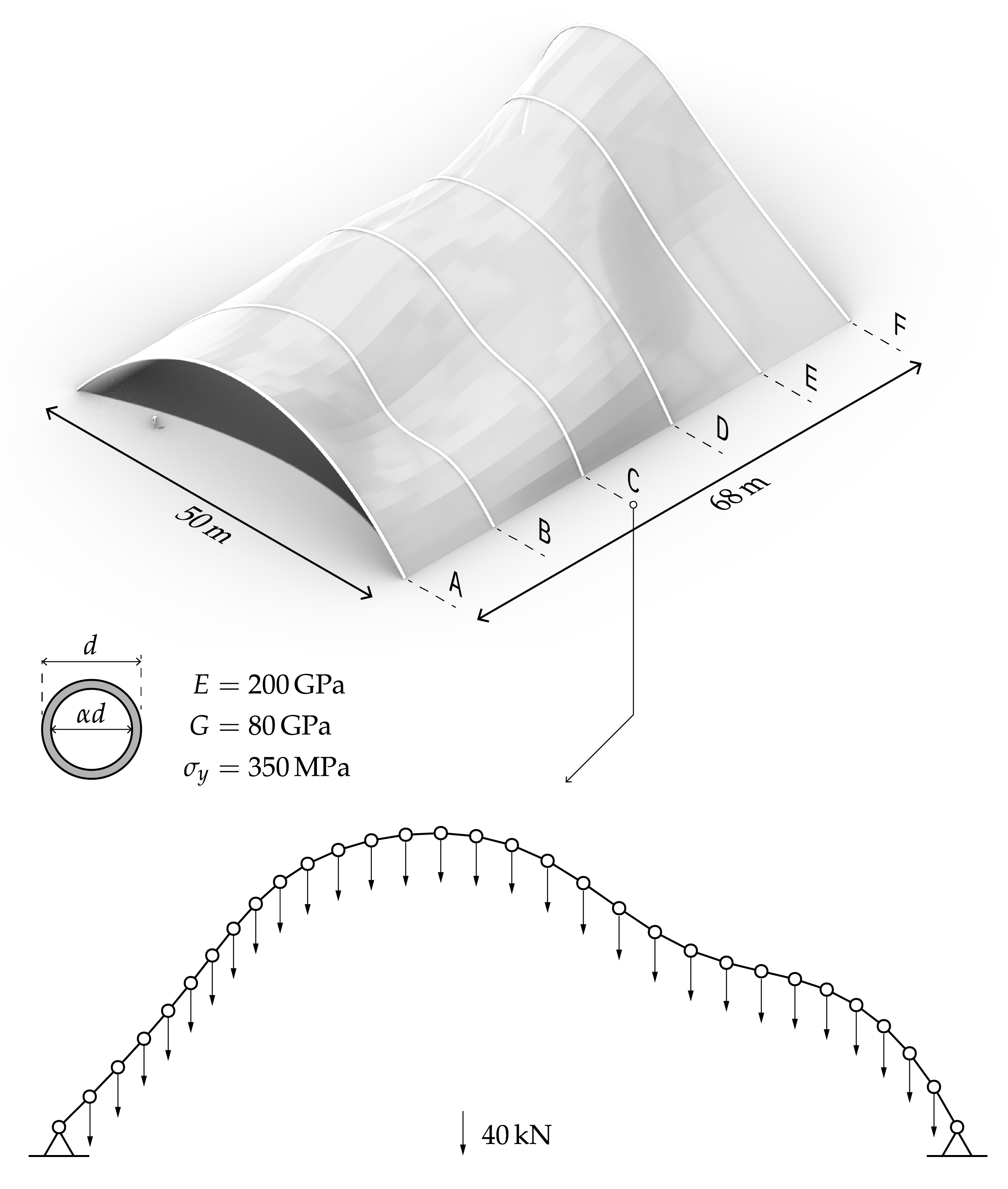}
  \caption{\label{fig:funicularize-1} A freeform spanning structure supported by six steel tubular frames. Each frame is discretized into 30 linear elements, pin supported at each end, and loaded at all free nodes.}
\end{figure}

In this example, we leverage the performance and composability of our framework for a multi-stage optimization process in the design of the supporting frames of a freeform spanning structure, shown in \cref{fig:funicularize-1}.
The structure consists of six independent planar frames made of steel hollow tube sections, discretized into 30 linear elements;
the ends of each frame are pin supported, and a downward load of \(40\text{kN}\) is applied to each free node.
Apart from Frame A, which follows a catenary curve, all other frames experience significant bending stresses due to their irregular curvature.
We assume that the geometry of these frames is fixed, and consider minimum volume sizing strategies of the frame members subject to stress, displacement, and constructability constraints.

\subsubsection{Single section sizing}\label{sec:funicularize-singlesec}
For procurement and fabrication simplicity, we consider a scenario in which all six frames must be constructed using a single tubular steel cross section.
Our design vector, \(x \in \R^2\), then consist of two values: \(d\), the exterior diameter of the section, and \(\alpha\), the proportional interior diameter (see \cref{fig:funicularize-1}).
The minimum volume problem is then:
\begin{align}
  \min_x \quad & f(x) = V = \sum_{i}^{n_e} A(d,\alpha)L_i \label{eq:min-volume-single-tube}\\
  \text{s.t.} \quad & \abs{d_{zi}} \leq d_{\max} \quad \forall \; i \in 1,\dots, n_n \nonumber\\
  & \abs{\sigma_i} \leq \sigma_{\max} \quad \forall \; i \in 1,\dots,n_e \nonumber
\end{align}

We set \(d_{\max} = 50\text{m} / 300 =  17\text{cm}\) and \(\sigma_{\max} = 350\text{MPa}\) as our vertical displacement and stress limits.
Unlike in previous examples, we model the structure using bending active frame elements with three translational and three rotational degrees of freedom at each node.
The additional section properties, such as the moment of inertia, \(I\), and torsional constant, \(J\), are functions of our design variables that must be computed along with the cross sectional area.
These properties are readily solved using known formulae for circular sections:
\begin{align*}
  A &= \frac{\pi d^2}{4} \left(1 - \alpha^2\right)\\
  I &= \frac{\pi d^4}{64} \left(1 - \alpha^4\right)\\
  J &= 2I
\end{align*}

\begin{figure}
  \centering
  \includegraphics{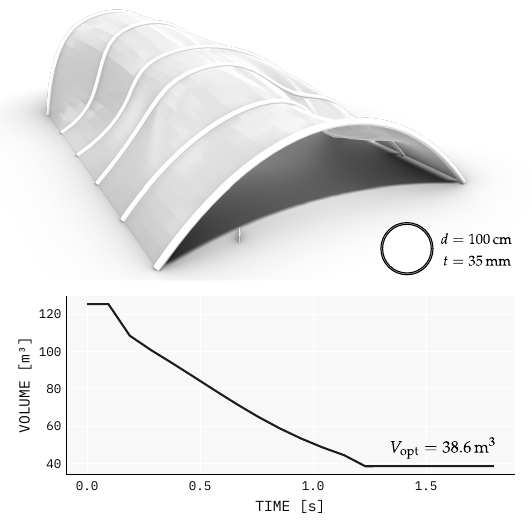}
  \caption{\label{fig:funicularize-2} Constrained minimum volume solution when using a single circular section. The solution is deflection dominated and seeks to maximize the bending stiffness of the frame through a large diameter.}
\end{figure}

The derivative of these properties with respect to our design variables are straightforward, and we do not define their explicit adjoint functions.
For stress constraints, we consider the combined effects of axial and flexural stresses through the summation:
\[
  \sigma = \frac{\abs{F}}{A} + \frac{\abs{M}}{S}
\]
where \(S = 2I/d\) is the section modulus, and \(M\) is the internal bending moment.
Setting our design variable values and bounds:
\begin{align*}
  0.1 \leq d_0 &= 0.75 \leq 1\\
  0.05 \leq \alpha_0 &= 0.5 \leq 0.98
\end{align*}
we solve \cref{eq:min-volume-single-tube} using our framework (\gbmma); the result is shown in \cref{fig:funicularize-2}.

Our gradient based approach finds a local minimum solution in less than two seconds, giving a single section structural volume of 38.6 m\textsuperscript{3}.
As the structure is dominated by the deflection constraint, the resulting solution maximizes the allowable section diameter to maximize flexural stiffness, at the cost of highly underutilized members, with a peak internal stress of 106 MPa.
Nonetheless, we show that our framework can be deployed to solve a common structural design problem, in which both mechanical constraints (stress, displacement) and physical constraints (fixed geometry, fabrication complexity) inform the feasible space of solutions.

\subsubsection{Spine reinforcement and grouped sizing}\label{sec:funicularize-multisec}
\begin{figure*}
  \centering
  \includegraphics[width = \textwidth]{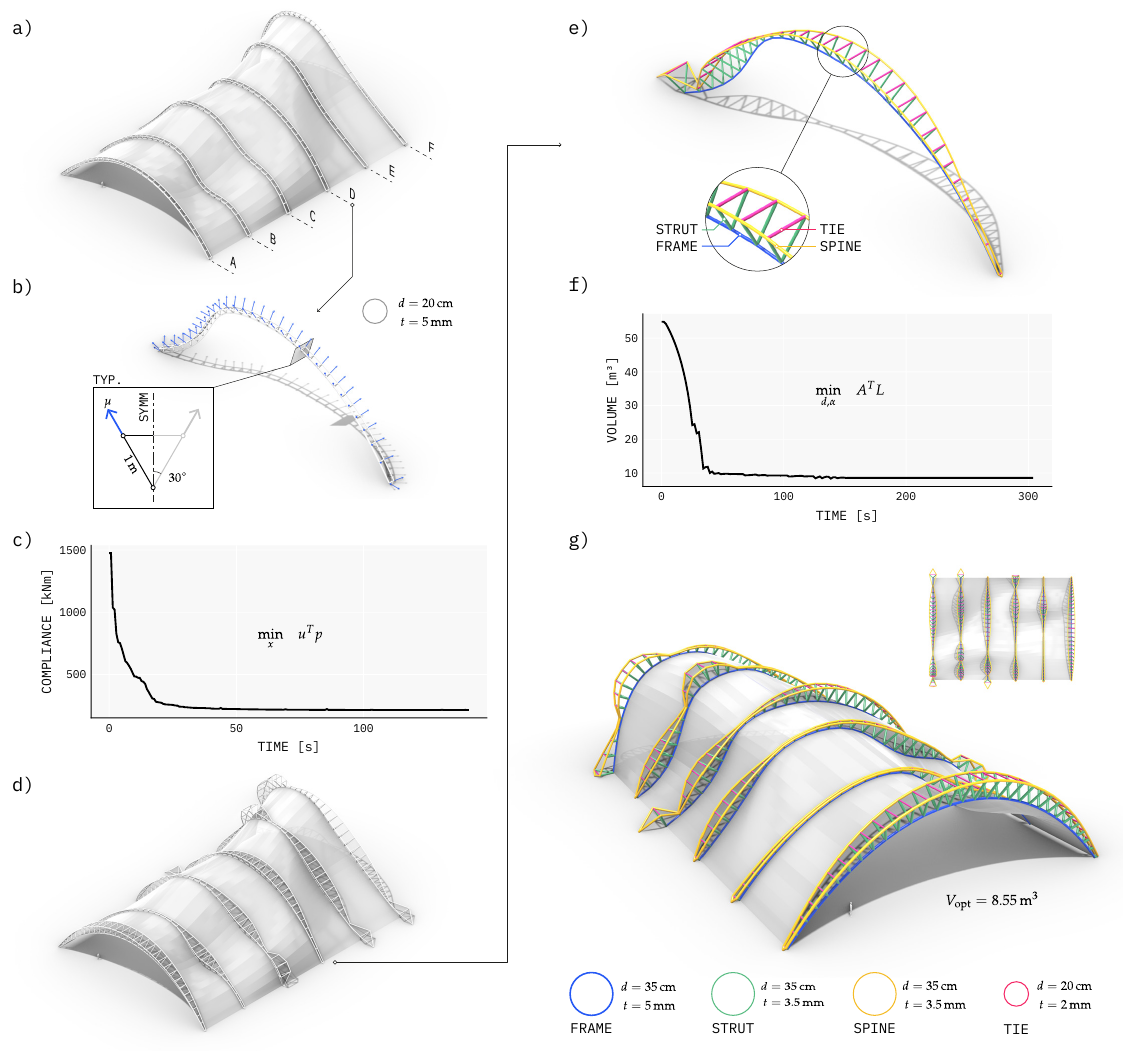}
  \caption{\label{fig:funicularize-3} A two-step optimization workflow for a minimum volume structure with fixed subset geometry. a) Augmenting initial frames with external Vierendeel truss spines. b) Defining spatial design variables of exterior nodes, constrained to perpendicular offsets from the initial frame. c) Unconstrained optimization to minimize structural compliance. d) Minimum compliance solution. e) Defining cross section groups for volume optimization. f) Constrained optimization to minimize the volume through the design of four unique hollow tube cross sections. g) Minimum volume solution and cross section geometries. Note that only one frame (Frame D) is highlighted in this figure, but all frames are optimized simultaneously at each step.}
\end{figure*}

We observed in \cref{sec:funicularize-singlesec} that deflection constraints govern the final solution, resulting in excessively large diameters assigned to the structural frames to provide sufficient stiffness.
Although the geometry of the frames must remain constant to support the freeform roof structure, we consider an alternative design strategy in which we stiffen the initial geometry with an external spine structure, and subsequently perform a minimum volume optimization.
Similar strategies have been deployed in practice, such as the international terminal at Waterloo station in London by Nicholas Grimshaw, and explored theoretically \cite{todisco2015Design,perez-sala2018Exploration}.
Intuitively, we seek to convert a deflection-dominated structural system into one that is either balanced or stress-dominated, enabling smaller cross sections and better material efficiency. 
Our approach, outlined in \cref{fig:funicularize-3}, consists of two sequential optimization problems: an unconstrained minimum compliance geometry problem to stiffen the augmented structure, and a constrained section design problem to minimize overall volume.

First, we augment the original structural system by adding a three dimensional Vierendeel truss system to the exterior of each frame (\cref{fig:funicularize-3}a).
A pair of perpendicular struts are extended from each frame node by a distance of one meter, angled outwards from the vertical axis by 30 degrees.
The ends of each strut pair are connected by a tie element, and the struts along each frame are connected together by a spine structure that is parallel to the orginal frame.
Two additional pinned supports are added to each side of the frame and connected to the spine ends to act as anchors.
Our design variables, outlined in \cref{fig:funicularize-3}b, are the changes in length of each strut pair from the initial one meter offset;
since the struts do not follow an orthogonal reference frame, we parameterize this variable by a scalar value \(\mu\) that is projected onto the original spine direction vector.
E.g., given a strut \(i\) whose orientation is represented by a vector \(v_i\), the \(x, y, z\) offset of the strut end is calculated by:
\begin{align*}
  \Delta x_i &= \mu_i v_{xi}\\
  \Delta y_i &= \mu_i v_{yi}\\
  \Delta z_i &= \mu_i v_{zi}
\end{align*}

We set the bounds of each variable to remain on the outside of the frame: \(-0.5 \leq \mu \leq 2.5\).
An additional spatial variable is defined for each anchor node, restricted to translation along the \(x\)-axis of the plane, bounded by up to a four meter extension away from the primary frame structure.
With 31 strut pairs and two anchors per frame, a total of 198 design variables are defined for this problem.
The cross section of all elements is held constant at \(20\,\text{cm}\) with a \(5\,\text{mm}\) wall thickness.

We solve for the optimal geometry of all six frames simultaneously using the L-BFGS algorithm \cite{liu1989Limited}; the objective value history and the solution geometry are shown in \cref{fig:funicularize-3}c and \cref{fig:funicularize-3}d.
The spines of each frame extend proportionally to regions of the original structure that experience large bending deformation, and generally seek to maintain a smooth parabolic pathway between the two primary supports.
This solution represents a minimum deformation local optimum given our problem parameters, and acts as the basis for constrained section design.

From here, we once again construct a direct section sizing problem to determine the minimum volume feasible solution.
Rather than a single section applied to all elements, as in \cref{sec:funicularize-singlesec}, we simultaneously design four unique sections corresponding to the augmented structural system: the main frame elements, the perpendicular strut elements, the parallel spine elements, and the connecting tie elements.
These section groups are highlighted in \cref{fig:funicularize-3}e.

Our design vector is then \(x \in \R^8\), corresponding to the \(d, \alpha\) parameters for each section family, with the same initial values and bounds:
\begin{align*}
  0.2 \leq &d_0 = 0.2 \leq 1\\
  0.01 \leq &\alpha_0 = 0.5 \leq 0.98
\end{align*}

Along with the same stress and displacement constraints defined in \cref{eq:min-volume-single-tube}, we define two additional geometric constraints to facilitate connections between different section types: (1) the strut diameter must be equal to or less than the diameter of the frame and spine elements, and (2) the tie diameter must be equal to or less than that of the spine elements.
These constraints ensure that a simple miter-and-weld connection method can be used in the fabrication of the structure.
We formalize the constrained optimization problem as:
\begin{align}
  \min_{d, \alpha} \quad & V = \transpose{A}L \label{eq:minvolume-multisec}\\
  \text{s.t.} \quad & \abs{d_{zi}} \leq d_{\max} \nonumber\\
  & \abs{\sigma_i} \leq \sigma_{\max} \nonumber\\
  &d_\text{strut} \leq \min\left(d_\text{frame}, d_\text{spine}\right) \nonumber\\
  &d_\text{tie} \leq d_\text{spine} \nonumber
\end{align}

The optimization history and final structural geometry are shown in \cref{fig:funicularize-3}f and \cref{fig:funicularize-3}g.
Three of the four section groups (frame, strut, and spine) are driven to the same final diameter, and the strut and spine sections also share the same wall thickness.
The tie element, experiencing the least amount of internal stress and providing minimal effect to overall structural stiffness, is driven to the smallest section size.
Critically, the final optimal volume is reduced from \(38.6\,\text{m}^3\) in the initial design to \(8.55\,\text{m}^3\) in the augmented structure.
Although this approach introduces additional complexity by way of increased element count and two additional section sizes, it enables a reduction of \(78\%\) in material from the original solution while meeting all structural constraints and enabling simple fabrication methods.

In this example, we showed how our framework enables rapid exploration and evaluation of different design strategies with varying objectives, constraints, and parameterizations:
we directly operate on cross section geometry to perform a direct optimal sizing of a structural design;
spatial design variables are projected along fixed vectors to perform restricted shape optimization;
cross section grouping and geometric constraints enable control over the complexity and constructability of the structure a priori when solving for minimum volume solutions.
Other strategies are readily accessible to explore in this framework: 
an increased allowance for complexity may allow for each frame segment to be sized independently; 
a series of different strut angles can be explored for the augmented structure and compared; 
the geometry of the frame itself may be allowed to change up to set limits.
The flexibility in parameterization and the inherent speed of gradient-based optimization enables such exploration, providing a crucial tool in the design stage of low volume structural systems.

\subsection{Minimum embodied carbon design}\label{sec:multimat}
So far, we have only considered minimum volume problems of a single material.
Such solutions have compounding benefits in the construction of building structures: reduced material cost, reduced transportation loads, and reduced forces on foundations.
Of increasing importance is the reduction in embodied carbon (EC), which is proportional to the quantity of material consumed.
However, if one seeks to directly minimize the EC of a given structure, an additional dimension beyond geometry (both global and local) must be considered---the choice of structural material itself.

Generally speaking, there is a positive correlation between the strength and stiffness of a material and its EC \cite{hammond2011Embodied}.
Low-carbon materials such as conventional or engineered wood come at the cost of lower stress limits and elastic modulii, requiring more volume to resist a given load; in contrast, high strength materials such as steel have inherently large embodied carbon per unit volume.
A comparison of the typical material properties of steel and glulam are shown in \cref{tab:steel-glulam-properties}.
Previous work in truss optimization has shown that the lowest embodied carbon solution typically combines the benefits of both ends of this spectrum \cite{ching2022Truss,stern2018Minimizing}.

\begin{figure}
  \centering
  \includegraphics{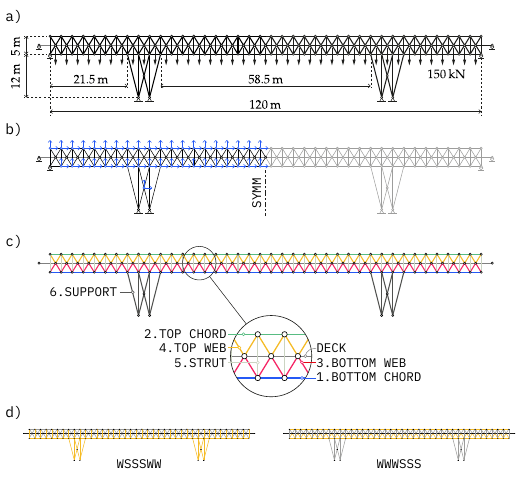}
  \caption{\label{fig:multimat-init} We seekthe minimum embodied carbon solution of a doubly cantilevered truss bridge using two materials. a) Initial structural design and boundary conditions. b) Spatial design variables include the \(x,y\) position of all free nodes, mirrored across the line of symmetry. c) We define six element groups to which we assign a single a material and cross section for optimization. d) two example initial structures with material assignments in order listed in c).}
\end{figure}

The key challenge in these multimaterial optimization strategies is the introduction of a combinatorial design variable: \emph{which} elements should be assigned \emph{which} material for the optimal design?
Naturally, the material properties for a given element in an optimization problem cannot be treated as continuous design variables.
They must rather be assigned and fixed before optimization, leading to a combinatorial expansion of the number of independent problems to solve.
Paired with inefficient gradient-free solvers, this process becomes untenable for structures of even moderate complexity.
In this example, we show how the speed of gradient-based optimization as well as the grouped sizing approach demonstrated in \cref{sec:spaceframe-rationalization} and \cref{sec:funicularize-multisec} can be deployed to exhaustively sample combinations of steel and glulam sections in a constrained minimium embodied carbon problem.

\begin{table}[]
  \caption{Properties of steel and glulam, where \(\text{ECC}\) is the Embodied Carbon Coefficient, \(\sigma_c\) and \(\sigma_t\) denote the compressive and tensile material strengths, and \(\rho\) is the material density.}
  \label{tab:steel-glulam-properties}
  \resizebox{\columnwidth}{!}{%
  \begin{tabular}{lcccc}
  \hline
         & \text{ECC} [kg CO\textsubscript{2}e / kg] & $\sigma_c$ [MPa] & $\sigma_t$ [MPa] & $\rho$ [kg / m\textsuperscript{3}] \\ \hline
  Steel  & 1.55                               & 350              & 350              & 7800                               \\
  Glulam & 0.512                              & 20.4             & 33               & 560                               
  \end{tabular}%
  }
\end{table}

Our problem formulation is outlined in \cref{fig:multimat-init}.
We seek to find the minimum embodied carbon solution to a symmetric doubly cantilevered truss bridge with a main span of \(56.5\,\text{m}\).
The structure supports a longitudinal deck sandwiched between the top and bottom chords, on which \(150\,\text{kN}\) point loads are applied; the deck remains static in our problem, and its embodied carbon is not considered in our objective function.

The \(x,y\) positions of all unsupported nodes (excluding deck) are set as our spatial design variables (\cref{fig:multimat-init}b), with bounds set to ensure that nodes do not cross over each other or generate elements of zero length.
For our cross sectional area variables, we categorize the structural elements into six groups (\cref{fig:multimat-init}c): (1) bottom chord, (2) top chord, (3) bottom web, (4) top web, (5) strut, and (6) support.
With two material choices, we have \(2^6=64\) possible material combinations to assign to the groups.

\begin{figure*}
  \centering
  \includegraphics[width = \textwidth]{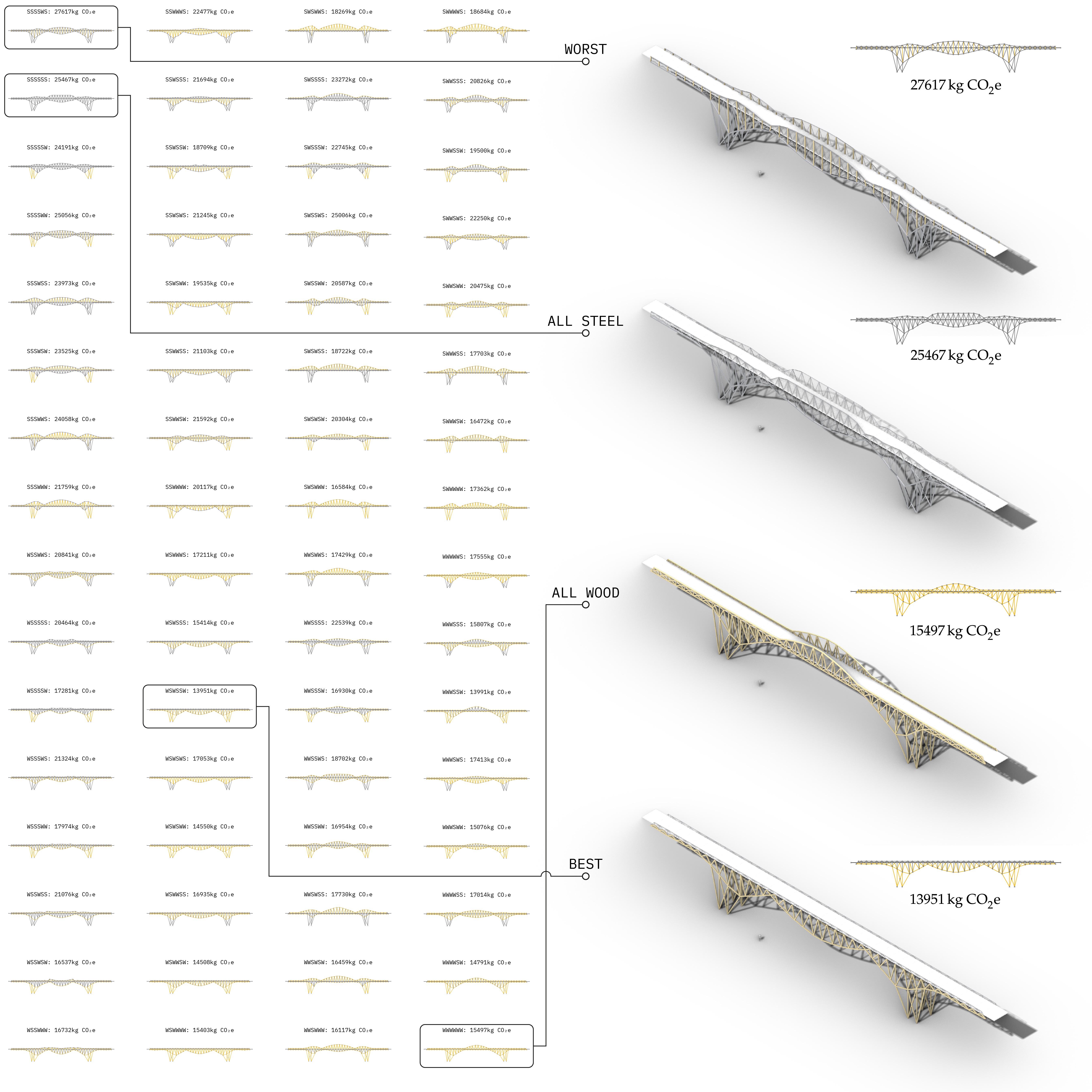}
  \caption{\label{fig:multimat-solutions} Minimum embodied carbon solutions for 64 material combinations. The overall optimal solution combines both steel and wood elements to an overall embodied carbon of \(13951\,\text{kg CO}_2\text{e}\), less than half of the worst optimal solution. Both single-material strategies result in optimal solutions that are neither the best or worst in the overall set of results.}
\end{figure*}

For each assignment combination, we define an ID by the material assigned to each group in the order listed in \cref{fig:multimat-init}c; for example, \texttt{WSSWWW} denotes a structure with a wood bottom chord, steel top chord, steel bottom web, wood top web, wood struts, and wood supports. 
The initial starting values and bounds are fixed for each material as:
\begin{align*}
  0.06 \leq A_{W0} &= 0.5 \leq 1.8\\
  0.001 \leq A_{S0} &= 0.034 \leq 0.2
\end{align*}
and with fixed material properties, defined in \cref{tab:steel-glulam-properties}.
The constrained optimization problem is formalized as:
\begin{align}
  \min_{x, A} \quad & \text{EC} = \text{ECC}_W \rho_W V_W + \text{ECC}_S  \rho_S V_S \label{eq:multimat-problem}\\
  \text{s.t.} \quad & |u_{\text{deck}}| \leq u_{\max} = 0.15 \nonumber\\
  & \sigma_{Wc} \leq \sigma_W \leq \sigma_{Wt} \nonumber\\
  & \sigma_{Sc} \leq \sigma_S \leq \sigma_{St} \nonumber
\end{align}
where \(\text{ECC}\) is the embodied carbon coefficient and \(\rho\) is the material density, defined in \cref{tab:steel-glulam-properties}.
The stress constraints are split by material: unlike steel, we assign different stress limits for glulam members under compression (\(\sigma_c\)) and tension (\(\sigma_t\)).
As before, we use the MMA optimization algorithm with a \(1 \times 10^{-6}\) stopping tolerance, and a time limit of two minutes for each run, giving an upper time limit of just over two hours.
An overview of all results is shown in \cref{fig:multimat-solutions}.

The total runtime of all 64 optimization problems is 4207s, averaging one minute and six seconds per problem.
As expected, material choice plays a significant role in environmental performance, with the worst performing design (\texttt{SSSSWS}) resulting in almost double the embodied carbon of the best material combination (\texttt{WSWSSW}).
The geometries of the optimal solutions also vary depending on material strategy, and do not solely rely on minimizing the lengths of all members, as seen in \cref{sec:spaceframe}.
This variation occurs primarily in the distribution of structural depth, which ranges from a conventional cantilever bridge typology with raised chords at both the supports and the central span (e.g., \texttt{SSSSSW}) to a single arch between supports (e.g., \texttt{WWWWWW}).
Neither monomaterial design strategy results in the best (or worst) embodied carbon solution, highlighting the importance of material strategies in low-carbon design.
Here, the inherent speed of gradient-based optimization plays a central role in evaluating such strategies in reasonable time; the total time for all 64 optimization solutions is less than that of a single run performed by \citet{stern2018Minimizing}.

\begin{figure}
  \centering
  \includegraphics{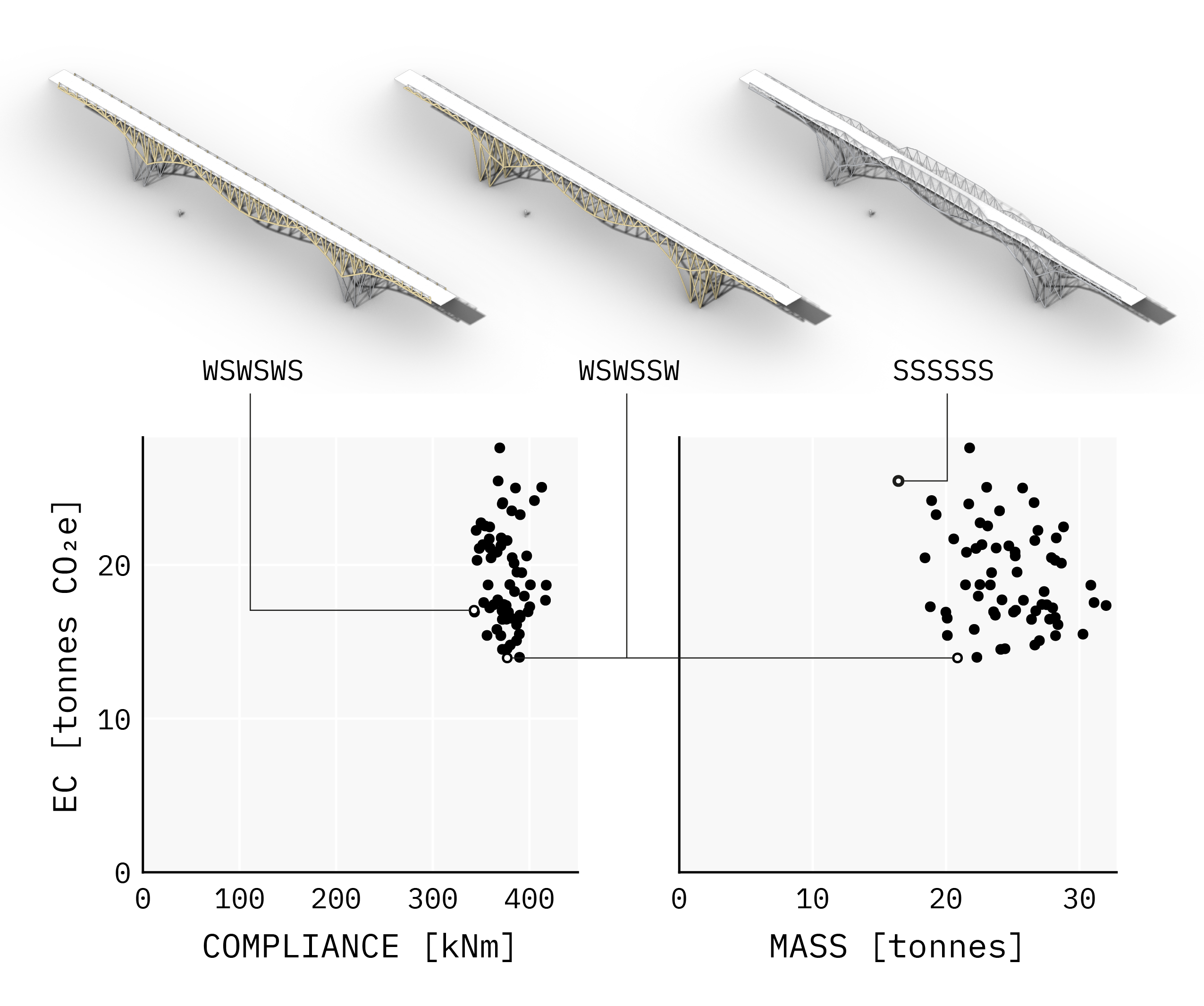}
  \caption{\label{fig:multimat-moo} Minimum embodied carbon solutions have no correlation with either structural compliance or overall mass. Our framework enables a direct minimization of important objectives rather than using proxy measures of performance.}
\end{figure}

We further emphasize the value of directly minimizing a desired objective function, rather than through proxy measures of performance, such as structural compliance.
In \cref{fig:multimat-moo}, we compare the optimal embodied carbon values of all 64 material combinations against two common objectives: structural compliance and total mass.
There is little correlation across objectives, and the emergence of a Pareto optimality front is observed in both plots.
Through our framework, we can directly minimize the enbodied carbon of a design while considering real engineering constraints such as anisotropic stress limits and global deflection.

\section{Conclusions} \label{sec:conclusion}
\subsection{Summary of contributions}
In this paper, we show how fast and scalable gradient-based optimization is possible for a significantly larger variety of structural design problems than previously thought feasible.
We achieve this through Automatic Differentiation, in which the complex computational graph of a typical structural analysis program is automatically generated to identify the topology of operations in a derivative calculation.
To actually use this graph efficiently when computing gradients, we derive the analytic derivatives of performance-critical intermediate functions and leverage the nature of backpropagation to allow for fast and memory-efficient accumulation of partial sensitivities.
A complete implementation of our framework in the Julia programming language is provided as an open-source repository for validation, application, and extension.

We test our framework on four different design problems to highlight the potential impact of gradient-based structural optimization in realistic secnarios.
In \cref{sec:warren} and \cref{sec:spaceframe}, we use our framework to directly minimize the structural volume of steel trusses subject to displacement and stress constraints.
We compare our method to four other optimization frameworks and show the benefits in speed, stability, and the quality of the overall solution.
In \cref{sec:funicularize} we show how a combination of unconstrained shape optimization and the direct sizing of tubular steel elements can be combined to explore efficient structural solutions to freeform architectural geometries.
In \cref{sec:multimat} we explore timber-steel hybrid structures to find minimum embodied carbon solutions for a multispan bridge structure, and show the importance of optimization speed when evaluating multiple starting conditions.

\subsection{Discussion and future work}
This paper demonstrates both the feasibility and capability of gradient-based optimization for realistic structural design problems, with impacts on material consumption, design complexity, and embodied carbon.
Although speed is generally the primary benefit of gradient-based methods, we note that in \cref{sec:warren} and \cref{sec:spaceframe}, the Genetic Algorithm (\gfga) converges at a much faster rate and results in an objective value close to our method.
One may argue that the effort spent in this paper may be of little practical value when evolutionary methods generally provide similar performance scores in similar timeframes.
However, as discussed in \cref{sec:spaceframe} we argue that there are implicit benefits to gradient-based methods, primarily: (1) a degree of design control, in which gradients trace smooth pathways through the design space towards a final geometry that also reflect a smooth transition from the original design, and (2) an underlying reflection of the true state of the structural system, in which patterns in force distributions are naturally reflected by the design variables.
Our framework acts as an available tool for designers in scenarios where these benefits are valued, and we suspect that a hybrid approach may result in a diverse range of interesting solutions. For example, initially using an evolutionary algorithm and then branching from multiple points along the optimization history to perform local gradient-based optimization.

There are numerous avenues for future work.
First, in terms of improvements to computational implementation, the benefits of parallelization, GPU computing, and additional memory management should be explored to maximize performance.
Although this paper focuses purely on reverse-mode AD, there are scenarios where forward-mode AD would theoretically have better performance, such as when the number of constraints greatly exceed the number of design variables, and would merit formal exploration.
Second, in terms of applicability in real design problems, we intend to investigate additional design constraints such as slenderness effects and buckling (both at the cross section and element scale), composite action, natural frequency requirements, etc.
For performance, some of these constraints may warrant explicit adjoint derivations.
Third, a general differentiable analysis framework enables complete interoperability with machine learning models, in which gradients can propagate between the weights of an ML network and the variables of a structural model.
This benefits model training, in which structural performance can be incorporated into the loss function,  and allows for new design parameterization, in which the output of a model can inform the variables of a structural design.

\subsection{Concluding remarks}
High-performance structural design should no longer be considered a desirable outcome, but a necessary one
in light of the ongoing climate crisis and a constantly growing global population.
Optimization has been a known pathway towards reaching high-performance solutions, but has long been constrained by the technical limitations in efficient computation at large scales.
Through the accessibility of gradients of complex design problems, this paper bridges a significant gap between theory and practice, enabling direct optimization of meaningful objectives for a wide range of structural systems.

\section*{Funding sources}
This research did not receive any specific grant from funding agencies in the public, commercial, or not-for-profit sectors.




\bibliographystyle{elsarticle-num-names} 
\bibliography{DiffAnalysis}

\begin{thebibliography}{64}
\expandafter\ifx\csname natexlab\endcsname\relax\def\natexlab#1{#1}\fi
\providecommand{\url}[1]{\texttt{#1}}
\providecommand{\href}[2]{#2}
\providecommand{\path}[1]{#1}
\providecommand{\DOIprefix}{doi:}
\providecommand{\ArXivprefix}{arXiv:}
\providecommand{\URLprefix}{URL: }
\providecommand{\Pubmedprefix}{pmid:}
\providecommand{\doi}[1]{\href{http://dx.doi.org/#1}{\path{#1}}}
\providecommand{\Pubmed}[1]{\href{pmid:#1}{\path{#1}}}
\providecommand{\bibinfo}[2]{#2}
\ifx\xfnm\relax \def\xfnm[#1]{\unskip,\space#1}\fi
\bibitem[{Bends{\o}e(1995)}]{bendsoe1995Optimization}
\bibinfo{author}{M.~P. Bends{\o}e}, \bibinfo{title}{Optimization of Structural Topology, Shape, and Material}, volume \bibinfo{volume}{414}, \bibinfo{publisher}{Springer}, \bibinfo{address}{Berlin}, \bibinfo{year}{1995}.
\bibitem[{Bends{\o}e and Sigmund(2003)}]{bendsoe2003Topology}
\bibinfo{author}{M.~P. Bends{\o}e}, \bibinfo{author}{O.~Sigmund}, \bibinfo{title}{Topology Optimization: Theory, Methods, and Applications}, \bibinfo{publisher}{Springer}, \bibinfo{address}{Berlin ; New York}, \bibinfo{year}{2003}.
\bibitem[{Council(2019)}]{worldgreenbuildingcouncil2019Bringinga}
\bibinfo{author}{W.~G.~B. Council}, \bibinfo{title}{Bringing Embodied Carbon Upfront}, \bibinfo{type}{Technical Report}, \bibinfo{year}{2019}.
\bibitem[{Michell(1904)}]{michell1904Limits}
\bibinfo{author}{A.~G.~M. Michell},
\newblock \bibinfo{title}{The limits of economy of material in frame-structures},
\newblock \bibinfo{journal}{The London, Edinburgh, and Dublin Philosophical Magazine and Journal of Science}  (\bibinfo{year}{1904}). \DOIprefix\doi{10.1080/14786440409463229}.
\bibitem[{Maxwell(1864)}]{maxwell1864Reciprocal}
\bibinfo{author}{J.~C. Maxwell},
\newblock \bibinfo{title}{On reciprocal figures and diagrams of forces},
\newblock \bibinfo{journal}{The London, Edinburgh, and Dublin Philosophical Magazine and Journal of Science}  (\bibinfo{year}{1864}). \DOIprefix\doi{10.1080/14786446408643663}.
\bibitem[{Mazurek et~al.(2011)Mazurek, Baker, and Tort}]{mazurek2011Geometrical}
\bibinfo{author}{A.~Mazurek}, \bibinfo{author}{W.~F. Baker}, \bibinfo{author}{C.~Tort},
\newblock \bibinfo{title}{Geometrical aspects of optimum truss like structures},
\newblock \bibinfo{journal}{Structural and Multidisciplinary Optimization} \bibinfo{volume}{43} (\bibinfo{year}{2011}) \bibinfo{pages}{231--242}. \DOIprefix\doi{10.1007/s00158-010-0559-x}.
\bibitem[{Baker et~al.(2013)Baker, Beghini, Mazurek, Carrion, and Beghini}]{baker2013Maxwell}
\bibinfo{author}{W.~F. Baker}, \bibinfo{author}{L.~L. Beghini}, \bibinfo{author}{A.~Mazurek}, \bibinfo{author}{J.~Carrion}, \bibinfo{author}{A.~Beghini},
\newblock \bibinfo{title}{Maxwell's reciprocal diagrams and discrete {{Michell}} frames},
\newblock \bibinfo{journal}{Structural and Multidisciplinary Optimization} \bibinfo{volume}{48} (\bibinfo{year}{2013}) \bibinfo{pages}{267--277}. \DOIprefix\doi{10.1007/s00158-013-0910-0}.
\bibitem[{Fang et~al.(2023)Fang, Brown, De~Wolf, and Mueller}]{fang2023Reducing}
\bibinfo{author}{D.~Fang}, \bibinfo{author}{N.~Brown}, \bibinfo{author}{C.~De~Wolf}, \bibinfo{author}{C.~Mueller},
\newblock \bibinfo{title}{Reducing embodied carbon in structural systems: {{A}} review of early-stage design strategies},
\newblock \bibinfo{journal}{Journal of Building Engineering} \bibinfo{volume}{76} (\bibinfo{year}{2023}) \bibinfo{pages}{107054}. \DOIprefix\doi{10.1016/j.jobe.2023.107054}.
\bibitem[{Barnett(1961)}]{barnett1961MinimumWeight}
\bibinfo{author}{R.~L. Barnett},
\newblock \bibinfo{title}{Minimum-{{Weight Design}} of {{Beams}} for {{Deflection}}},
\newblock \bibinfo{journal}{Journal of the Engineering Mechanics Division} \bibinfo{volume}{87} (\bibinfo{year}{1961}) \bibinfo{pages}{75--109}. \DOIprefix\doi{10.1061/JMCEA3.0000208}.
\bibitem[{Shield and Prager(1970)}]{shield1970Optimal}
\bibinfo{author}{R.~T. Shield}, \bibinfo{author}{W.~Prager},
\newblock \bibinfo{title}{Optimal structural design for given deflection},
\newblock \bibinfo{journal}{Zeitschrift f{\"u}r angewandte Mathematik und Physik ZAMP} \bibinfo{volume}{21} (\bibinfo{year}{1970}) \bibinfo{pages}{513--523}. \DOIprefix\doi{10.1007/BF01587681}.
\bibitem[{Prager and Rozvany(1977)}]{prager1977Optimala}
\bibinfo{author}{W.~Prager}, \bibinfo{author}{G.~I.~N. Rozvany},
\newblock \bibinfo{title}{Optimal {{Layout}} of {{Grillages}}},
\newblock \bibinfo{journal}{Journal of Structural Mechanics} \bibinfo{volume}{5} (\bibinfo{year}{1977}) \bibinfo{pages}{1--18}. \DOIprefix\doi{10.1080/03601217708907301}.
\bibitem[{Prager(1970)}]{prager1970Optimization}
\bibinfo{author}{W.~Prager},
\newblock \bibinfo{title}{Optimization of structural design},
\newblock \bibinfo{journal}{Journal of Optimization Theory and Applications} \bibinfo{volume}{6} (\bibinfo{year}{1970}) \bibinfo{pages}{1--21}. \DOIprefix\doi{10.1007/BF00927037}.
\bibitem[{Kirsch and Taye(1986)}]{kirsch1986Optimal}
\bibinfo{author}{U.~Kirsch}, \bibinfo{author}{S.~Taye},
\newblock \bibinfo{title}{On optimal topology of grillage structures},
\newblock \bibinfo{journal}{Engineering with Computers} \bibinfo{volume}{1} (\bibinfo{year}{1986}) \bibinfo{pages}{229--243}. \DOIprefix\doi{10.1007/BF01200139}.
\bibitem[{Bolbotowski et~al.(2018)Bolbotowski, He, and Gilbert}]{bolbotowski2018Design}
\bibinfo{author}{K.~Bolbotowski}, \bibinfo{author}{L.~He}, \bibinfo{author}{M.~Gilbert},
\newblock \bibinfo{title}{Design of optimum grillages using layout optimization},
\newblock \bibinfo{journal}{Structural and Multidisciplinary Optimization} \bibinfo{volume}{58} (\bibinfo{year}{2018}) \bibinfo{pages}{851--868}. \DOIprefix\doi{10.1007/s00158-018-1930-6}.
\bibitem[{Whiteley et~al.(2023)Whiteley, Liew, He, and Gilbert}]{whiteley2023Engineering}
\bibinfo{author}{J.~Whiteley}, \bibinfo{author}{A.~Liew}, \bibinfo{author}{L.~He}, \bibinfo{author}{M.~Gilbert},
\newblock \bibinfo{title}{Engineering design of optimized reinforced concrete floor grillages},
\newblock \bibinfo{journal}{Structures} \bibinfo{volume}{51} (\bibinfo{year}{2023}) \bibinfo{pages}{1292--1304}. \DOIprefix\doi{10.1016/j.istruc.2023.03.116}.
\bibitem[{Rozvany et~al.(1995)Rozvany, Bendsoe, and Kirsch}]{rozvany1995Layout}
\bibinfo{author}{G.~I.~N. Rozvany}, \bibinfo{author}{M.~P. Bendsoe}, \bibinfo{author}{U.~Kirsch},
\newblock \bibinfo{title}{Layout {{Optimization}} of {{Structures}}},
\newblock \bibinfo{journal}{Applied Mechanics Reviews} \bibinfo{volume}{48} (\bibinfo{year}{1995}) \bibinfo{pages}{41--119}. \DOIprefix\doi{10.1115/1.3005097}.
\bibitem[{Mei and Wang(2021)}]{mei2021Structural}
\bibinfo{author}{L.~Mei}, \bibinfo{author}{Q.~Wang},
\newblock \bibinfo{title}{Structural {{Optimization}} in {{Civil Engineering}}: {{A Literature Review}}},
\newblock \bibinfo{journal}{Buildings} \bibinfo{volume}{11} (\bibinfo{year}{2021}) \bibinfo{pages}{66}. \DOIprefix\doi{10.3390/buildings11020066}.
\bibitem[{Ismail et~al.(2021)Ismail, Mayencourt, and Mueller}]{ismail2021Shaped}
\bibinfo{author}{M.~A. Ismail}, \bibinfo{author}{P.~L. Mayencourt}, \bibinfo{author}{C.~T. Mueller},
\newblock \bibinfo{title}{Shaped beams: Unlocking new geometry for efficient structures},
\newblock \bibinfo{journal}{Architecture, Structures and Construction} \bibinfo{volume}{1} (\bibinfo{year}{2021}) \bibinfo{pages}{37--52}. \DOIprefix\doi{10.1007/s44150-021-00003-y}.
\bibitem[{Mayencourt and Mueller(2019)}]{mayencourt2019Structural}
\bibinfo{author}{P.~Mayencourt}, \bibinfo{author}{C.~Mueller},
\newblock \bibinfo{title}{Structural {{Optimization}} of {{Cross-laminated Timber Panels}} in {{One-way Bending}}},
\newblock \bibinfo{journal}{Structures} \bibinfo{volume}{18} (\bibinfo{year}{2019}) \bibinfo{pages}{48--59}. \DOIprefix\doi{10.1016/j.istruc.2018.12.009}.
\bibitem[{Preisinger et~al.(2011)Preisinger, Vierlinger, Hofmann, and Bollinger}]{preisinger2011Evolutionary}
\bibinfo{author}{C.~Preisinger}, \bibinfo{author}{R.~Vierlinger}, \bibinfo{author}{A.~Hofmann}, \bibinfo{author}{K.~Bollinger}, \bibinfo{title}{Evolutionary {{Structural Optimization Revisited}}}, \bibinfo{year}{2011}. \DOIprefix\doi{10.13140/RG.2.2.11221.27361}.
\bibitem[{Vierlinger and Hofmann(2013)}]{vierlinger2013Framework}
\bibinfo{author}{R.~Vierlinger}, \bibinfo{author}{A.~Hofmann}, \bibinfo{title}{A {{Framework}} for {{Flexible Search}} and {{Optimization}} in {{Parametric Design}}}, \bibinfo{year}{2013}. \DOIprefix\doi{10.13140/RG.2.1.1516.8727}.
\bibitem[{Bletzinger and Maute(1997)}]{bletzinger1997Generalized}
\bibinfo{author}{K.-U. Bletzinger}, \bibinfo{author}{K.~Maute},
\newblock \bibinfo{title}{Towards {{Generalized Shape}} and {{Topology Optimization}}},
\newblock \bibinfo{journal}{Engineering Optimization} \bibinfo{volume}{29} (\bibinfo{year}{1997}) \bibinfo{pages}{201--216}. \DOIprefix\doi{10.1080/03052159708940993}.
\bibitem[{Bletzinger and Ramm(2001)}]{bletzinger2001Structural}
\bibinfo{author}{K.-U. Bletzinger}, \bibinfo{author}{E.~Ramm},
\newblock \bibinfo{title}{Structural optimization and form finding of light weight structures},
\newblock \bibinfo{journal}{Computers \& Structures} \bibinfo{volume}{79} (\bibinfo{year}{2001}) \bibinfo{pages}{2053--2062}.
\bibitem[{Powell(1994)}]{powell1994Direct}
\bibinfo{author}{M.~J.~D. Powell},
\newblock \bibinfo{title}{A {{Direct Search Optimization Method That Models}} the {{Objective}} and {{Constraint Functions}} by {{Linear Interpolation}}},
\newblock in: \bibinfo{editor}{S.~Gomez}, \bibinfo{editor}{J.-P. Hennart} (Eds.), \bibinfo{booktitle}{Advances in {{Optimization}} and {{Numerical Analysis}}}, \bibinfo{publisher}{Springer Netherlands}, \bibinfo{address}{Dordrecht}, \bibinfo{year}{1994}, pp. \bibinfo{pages}{51--67}. \DOIprefix\doi{10.1007/978-94-015-8330-5_4}.
\bibitem[{Powell(2009)}]{powell2009BOBYQA}
\bibinfo{author}{M.~J.~D. Powell},
\newblock \bibinfo{title}{The {{BOBYQA}} algorithm for bound constrained optimization without derivatives}  (\bibinfo{year}{2009}).
\bibitem[{Kaveh(2017)}]{kaveh2017Advances}
\bibinfo{author}{A.~Kaveh}, \bibinfo{title}{Advances in {{Metaheuristic Algorithms}} for {{Optimal Design}} of {{Structures}}}, \bibinfo{publisher}{Springer International Publishing}, \bibinfo{address}{Cham}, \bibinfo{year}{2017}. \DOIprefix\doi{10.1007/978-3-319-46173-1}.
\bibitem[{Kicinger et~al.(2005)Kicinger, Arciszewski, and DeJong}]{kicinger2005Evolutionary}
\bibinfo{author}{R.~Kicinger}, \bibinfo{author}{T.~Arciszewski}, \bibinfo{author}{K.~DeJong},
\newblock \bibinfo{title}{Evolutionary {{Design}} of {{Steel Structures}} in {{Tall Buildings}}},
\newblock \bibinfo{journal}{Journal of Computing in Civil Engineering} \bibinfo{volume}{19} (\bibinfo{year}{2005}) \bibinfo{pages}{223--238}. \DOIprefix\doi{10.1061/(ASCE)0887-3801(2005)19:3(223)}.
\bibitem[{Xie and Steven(1993)}]{xie1993Simple}
\bibinfo{author}{Y.~Xie}, \bibinfo{author}{G.~Steven},
\newblock \bibinfo{title}{A simple evolutionary procedure for structural optimization},
\newblock \bibinfo{journal}{Computers \& Structures} \bibinfo{volume}{49} (\bibinfo{year}{1993}) \bibinfo{pages}{885--896}. \DOIprefix\doi{10.1016/0045-7949(93)90035-C}.
\bibitem[{Munk et~al.(2015)Munk, Vio, and Steven}]{munk2015Topology}
\bibinfo{author}{D.~J. Munk}, \bibinfo{author}{G.~A. Vio}, \bibinfo{author}{G.~P. Steven},
\newblock \bibinfo{title}{Topology and shape optimization methods using evolutionary algorithms: A review},
\newblock \bibinfo{journal}{Structural and Multidisciplinary Optimization} \bibinfo{volume}{52} (\bibinfo{year}{2015}) \bibinfo{pages}{613--631}. \DOIprefix\doi{10.1007/s00158-015-1261-9}.
\bibitem[{Griewank(1991)}]{griewank1991Automatic}
\bibinfo{author}{A.~Griewank},
\newblock \bibinfo{title}{Automatic differentiaion of algorithms: {{Theory}}, implementation and application},
\newblock \bibinfo{journal}{Siam} \bibinfo{volume}{1991} (\bibinfo{year}{1991}).
\bibitem[{Griewank and Walther(2008)}]{griewank2008Evaluating}
\bibinfo{author}{A.~Griewank}, \bibinfo{author}{A.~Walther}, \bibinfo{title}{Evaluating {{Derivatives}}: {{Principles}} and {{Techniques}} of {{Algorithmic Differentiation}}, {{Second Edition}}}, \bibinfo{edition}{second} ed., \bibinfo{publisher}{{Society for Industrial and Applied Mathematics}}, \bibinfo{year}{2008}. \DOIprefix\doi{10.1137/1.9780898717761}.
\bibitem[{Werbos(1990)}]{werbos1990Backpropagation}
\bibinfo{author}{P.~Werbos},
\newblock \bibinfo{title}{Backpropagation through time: What it does and how to do it},
\newblock \bibinfo{journal}{Proceedings of the IEEE} \bibinfo{volume}{78} (\bibinfo{year}{1990}) \bibinfo{pages}{1550--1560}. \DOIprefix\doi{10.1109/5.58337}.
\bibitem[{Margossian(2019)}]{margossian2019Review}
\bibinfo{author}{C.~C. Margossian},
\newblock \bibinfo{title}{A {{Review}} of automatic differentiation and its efficient implementation},
\newblock \bibinfo{journal}{Wiley Interdisciplinary Reviews: Data Mining and Knowledge Discovery}  (\bibinfo{year}{2019}) \bibinfo{pages}{e1305}. \DOIprefix\doi{10/gf495m}. \href{http://arxiv.org/abs/1811.05031}{{\tt arXiv:1811.05031}}.
\bibitem[{Hu et~al.(2020)Hu, Anderson, Li, Sun, Carr, {Ragan-Kelley}, and Durand}]{hu2020DiffTaichi}
\bibinfo{author}{Y.~Hu}, \bibinfo{author}{L.~Anderson}, \bibinfo{author}{T.-M. Li}, \bibinfo{author}{Q.~Sun}, \bibinfo{author}{N.~Carr}, \bibinfo{author}{J.~{Ragan-Kelley}}, \bibinfo{author}{F.~Durand}, \bibinfo{title}{{{DiffTaichi}}: {{Differentiable Programming}} for {{Physical Simulation}}}, \bibinfo{year}{2020}. \DOIprefix\doi{10.48550/arXiv.1910.00935}. \href{http://arxiv.org/abs/1910.00935}{{\tt arXiv:1910.00935}}.
\bibitem[{Schek(1974)}]{schek1974Force}
\bibinfo{author}{H.~J. Schek},
\newblock \bibinfo{title}{The force density method for form finding and computation of general networks},
\newblock \bibinfo{journal}{Computer Methods in Applied Mechanics and Engineering} \bibinfo{volume}{3} (\bibinfo{year}{1974}) \bibinfo{pages}{115--134}. \DOIprefix\doi{10.1016/0045-7825(74)90045-0}.
\bibitem[{Cuvilliers et~al.(2016)Cuvilliers, Danhaive, and Mueller}]{cuvilliers2016Gradientbased}
\bibinfo{author}{P.~Cuvilliers}, \bibinfo{author}{R.~Danhaive}, \bibinfo{author}{C.~Mueller},
\newblock \bibinfo{title}{Gradient-based optimization of closest-fit funicular structures}  (\bibinfo{year}{2016}) \bibinfo{pages}{10}.
\bibitem[{Pastrana and Oktay(2024)}]{pastrana2024Arpastrana}
\bibinfo{author}{R.~Pastrana}, \bibinfo{author}{D.~Oktay}, \bibinfo{title}{Arpastrana/jax\_fdm: {{Release}} v0.8.4}, \bibinfo{howpublished}{Zenodo}, \bibinfo{year}{2024}. \DOIprefix\doi{10.5281/zenodo.11164249}.
\bibitem[{Burke et~al.(2023)Burke, Lee, Echelman, Feldman, and Mueller}]{burke2023FDMremote}
\bibinfo{author}{A.~Burke}, \bibinfo{author}{K.~Lee}, \bibinfo{author}{J.~Echelman}, \bibinfo{author}{D.~Feldman}, \bibinfo{author}{C.~Mueller},
\newblock \bibinfo{title}{{{FDMremote}}: {{Interactive}} inverse design of tensile structures with differentiable {{FDM}}},
\newblock \bibinfo{journal}{Proceedings of IASS Annual Symposia} \bibinfo{volume}{2023} (\bibinfo{year}{2023}) \bibinfo{pages}{1--12}.
\bibitem[{Ole~Ohlbrock and Schwartz(2016)}]{oleohlbrock2016Combinatorial}
\bibinfo{author}{P.~Ole~Ohlbrock}, \bibinfo{author}{J.~Schwartz},
\newblock \bibinfo{title}{Combinatorial equilibrium modeling},
\newblock \bibinfo{journal}{International Journal of Space Structures} \bibinfo{volume}{31} (\bibinfo{year}{2016}) \bibinfo{pages}{177--189}. \DOIprefix\doi{10.1177/0266351116660799}.
\bibitem[{Pastrana et~al.(2022)Pastrana, Ohlbrock, Oberbichler, D'Acunto, and Parascho}]{pastrana2022Constrained}
\bibinfo{author}{R.~Pastrana}, \bibinfo{author}{P.~O. Ohlbrock}, \bibinfo{author}{T.~Oberbichler}, \bibinfo{author}{P.~D'Acunto}, \bibinfo{author}{S.~Parascho}, \bibinfo{title}{Constrained {{Form-Finding}} of {{Tension-Compression Structures}} using {{Automatic Differentiation}}}, \bibinfo{year}{2022}. \href{http://arxiv.org/abs/2111.02607}{{\tt arXiv:2111.02607}}.
\bibitem[{Wu(2023)}]{wu2023Framework}
\bibinfo{author}{G.~Wu},
\newblock \bibinfo{title}{A framework for structural shape optimization based on automatic differentiation, the adjoint method and accelerated linear algebra},
\newblock \bibinfo{journal}{Structural and Multidisciplinary Optimization} \bibinfo{volume}{66} (\bibinfo{year}{2023}) \bibinfo{pages}{151}. \DOIprefix\doi{10.1007/s00158-023-03601-0}.
\bibitem[{Kassimali(2012)}]{kassimali2012Matrix}
\bibinfo{author}{A.~Kassimali}, \bibinfo{title}{Matrix Analysis of Structures}, \bibinfo{edition}{2nd ed} ed., \bibinfo{publisher}{Cengage Learning}, \bibinfo{address}{Australia ; Stamford, CT}, \bibinfo{year}{2012}.
\bibitem[{Livesley(1956)}]{livesley1956Automatic}
\bibinfo{author}{R.~K. Livesley},
\newblock \bibinfo{title}{The automatic design of structural frames},
\newblock \bibinfo{journal}{The Quarterly Journal of Mechanics and Applied Mathematics} \bibinfo{volume}{9} (\bibinfo{year}{1956}) \bibinfo{pages}{257--278}. \DOIprefix\doi{10.1093/qjmam/9.3.257}.
\bibitem[{Turner et~al.(1956)Turner, Clough, Martin, and Topp}]{turner1956Stiffness}
\bibinfo{author}{M.~J. Turner}, \bibinfo{author}{R.~W. Clough}, \bibinfo{author}{H.~C. Martin}, \bibinfo{author}{L.~J. Topp},
\newblock \bibinfo{title}{Stiffness and {{Deflection Analysis}} of {{Complex Structures}}},
\newblock \bibinfo{journal}{Journal of the Aeronautical Sciences} \bibinfo{volume}{23} (\bibinfo{year}{1956}) \bibinfo{pages}{805--823}. \DOIprefix\doi{10.2514/8.3664}.
\bibitem[{Laue et~al.(2020)Laue, Mitterreiter, and Giesen}]{laue2020Simple}
\bibinfo{author}{S.~Laue}, \bibinfo{author}{M.~Mitterreiter}, \bibinfo{author}{J.~Giesen},
\newblock \bibinfo{title}{A {{Simple}} and {{Efficient Tensor Calculus}}},
\newblock \bibinfo{journal}{Proceedings of the AAAI Conference on Artificial Intelligence} \bibinfo{volume}{34} (\bibinfo{year}{2020}) \bibinfo{pages}{4527--4534}. \DOIprefix\doi{10.1609/aaai.v34i04.5881}.
\bibitem[{Martins and Ning(2021)}]{martins2021Engineering}
\bibinfo{author}{J.~R. R.~A. Martins}, \bibinfo{author}{A.~Ning}, \bibinfo{title}{Engineering {{Design Optimization}}}, \bibinfo{edition}{1} ed., \bibinfo{publisher}{Cambridge University Press}, \bibinfo{year}{2021}. \DOIprefix\doi{10.1017/9781108980647}.
\bibitem[{Giles(2008)}]{giles2008Extended}
\bibinfo{author}{M.~Giles}, \bibinfo{title}{An Extended Collection of Matrix Derivative Results for Forward and Reverse Mode Algorithmic Differentiation}, \bibinfo{type}{Technical Report} \bibinfo{number}{08/01}, Oxford University Computing Laboratory, \bibinfo{year}{2008}.
\bibitem[{Cichocki et~al.(2009)Cichocki, Zdunek, Phan, and Amari}]{cichocki2009Introduction}
\bibinfo{author}{A.~Cichocki}, \bibinfo{author}{R.~Zdunek}, \bibinfo{author}{A.~H. Phan}, \bibinfo{author}{S.-I. Amari},
\newblock \bibinfo{title}{Introduction -- {{Problem Statements}} and {{Models}}},
\newblock in: \bibinfo{booktitle}{Nonnegative {{Matrix}} and {{Tensor Factorizations}}}, \bibinfo{publisher}{John Wiley \& Sons, Ltd}, \bibinfo{year}{2009}, pp. \bibinfo{pages}{1--80}. \DOIprefix\doi{10.1002/9780470747278.ch1}.
\bibitem[{Davis(2004)}]{davis2004Algorithm}
\bibinfo{author}{T.~A. Davis},
\newblock \bibinfo{title}{Algorithm 832: {{UMFPACK V4}}.3---an unsymmetric-pattern multifrontal method},
\newblock \bibinfo{journal}{ACM Transactions on Mathematical Software} \bibinfo{volume}{30} (\bibinfo{year}{2004}) \bibinfo{pages}{196--199}. \DOIprefix\doi{10.1145/992200.992206}.
\bibitem[{Johnson(2021)}]{johnson2021Notes}
\bibinfo{author}{S.~G. Johnson}, \bibinfo{title}{Notes on {{Adjoint Methods}} for 18.335}, \bibinfo{type}{Technical Report}, Massachusetts Institute of Technology, \bibinfo{year}{2021}.
\bibitem[{Lee(2024)}]{lee2024Asapa}
\bibinfo{author}{K.~J. Lee}, \bibinfo{title}{Asap.jl}, \bibinfo{howpublished}{Zenodo}, \bibinfo{year}{2024}. \DOIprefix\doi{10.5281/zenodo.10805075}.
\bibitem[{Innes(2019)}]{innes2019Don}
\bibinfo{author}{M.~Innes}, \bibinfo{title}{Don't {{Unroll Adjoint}}: {{Differentiating SSA-Form Programs}}}, \bibinfo{year}{2019}. \DOIprefix\doi{10.48550/arXiv.1810.07951}. \href{http://arxiv.org/abs/1810.07951}{{\tt arXiv:1810.07951}}.
\bibitem[{White et~al.(2024)White, Abbott, Zgubic, Revels, Robinson, Arslan, Widmann, Schaub, Ma, Tebbutt, Axen, Rackauckas, Vertechi, BSnelling, Fischer, {st--}, Horikawa, Cottier, Ranocha, Bhagavan, Schmitz, Sajko, Besan{\c c}on, Marco, Jutho, Dalle, Chen, Anthony~Blaom, Schulz, and Haessig}]{white2024JuliaDiff}
\bibinfo{author}{F.~White}, \bibinfo{author}{M.~Abbott}, \bibinfo{author}{M.~Zgubic}, \bibinfo{author}{J.~Revels}, \bibinfo{author}{N.~Robinson}, \bibinfo{author}{A.~Arslan}, \bibinfo{author}{D.~Widmann}, \bibinfo{author}{S.~D. Schaub}, \bibinfo{author}{Y.~Ma}, \bibinfo{author}{W.~Tebbutt}, \bibinfo{author}{S.~Axen}, \bibinfo{author}{C.~Rackauckas}, \bibinfo{author}{P.~Vertechi}, \bibinfo{author}{BSnelling}, \bibinfo{author}{K.~Fischer}, \bibinfo{author}{{st--}}, \bibinfo{author}{Y.~Horikawa}, \bibinfo{author}{B.~Cottier}, \bibinfo{author}{H.~Ranocha}, \bibinfo{author}{S.~Bhagavan}, \bibinfo{author}{N.~Schmitz}, \bibinfo{author}{N.~Sajko}, \bibinfo{author}{M.~Besan{\c c}on}, \bibinfo{author}{Marco}, \bibinfo{author}{Jutho}, \bibinfo{author}{G.~Dalle}, \bibinfo{author}{B.~Chen}, \bibinfo{author}{P.~Anthony~Blaom}, \bibinfo{author}{O.~Schulz}, \bibinfo{author}{P.~Haessig}, \bibinfo{title}{{{JuliaDiff}}/{{ChainRulesCore}}.jl: V1.24.0}, \bibinfo{howpublished}{Zenodo}, \bibinfo{year}{2024}. \DOIprefix\doi{10.5281/zenodo.11406387}.
\bibitem[{Tarek(2023)}]{tarek2023Nonconvex}
\bibinfo{author}{M.~Tarek}, \bibinfo{title}{Nonconvex.Jl: {{A Comprehensive Julia Package}} for {{Non-Convex Optimization}}}, \bibinfo{year}{2023}. \DOIprefix\doi{10.13140/RG.2.2.36120.37121}.
\bibitem[{Lee(2024)}]{Lee_DiffAnalysis_AIC2024}
\bibinfo{author}{K.~J. Lee}, \bibinfo{title}{{{DiffAnalysis}}\_{{AIC2024}}}, \bibinfo{howpublished}{\url{https://github.com/keithjlee/DiffAnalysis_AutomationInConstruction_2024_CODE}}, \bibinfo{year}{2024}.
\bibitem[{Powell(1998)}]{powell1998Direct}
\bibinfo{author}{M.~J.~D. Powell},
\newblock \bibinfo{title}{Direct search algorithms for optimization calculations},
\newblock \bibinfo{journal}{Acta Numerica} \bibinfo{volume}{7} (\bibinfo{year}{1998}) \bibinfo{pages}{287--336}. \DOIprefix\doi{10.1017/S0962492900002841}.
\bibitem[{Golberg(1989)}]{golberg1989genetic}
\bibinfo{author}{D.~E. Golberg},
\newblock \bibinfo{title}{Genetic algorithms in search, optimization, and machine learning},
\newblock \bibinfo{journal}{Addion wesley} \bibinfo{volume}{1989} (\bibinfo{year}{1989}) \bibinfo{pages}{36}.
\bibitem[{Svanberg(1987)}]{svanberg1987Method}
\bibinfo{author}{K.~Svanberg},
\newblock \bibinfo{title}{The method of moving asymptotes---a new method for structural optimization},
\newblock \bibinfo{journal}{International Journal for Numerical Methods in Engineering} \bibinfo{volume}{24} (\bibinfo{year}{1987}) \bibinfo{pages}{359--373}. \DOIprefix\doi{10.1002/nme.1620240207}.
\bibitem[{Todisco et~al.(2015)Todisco, Fivet, {Corres-Peiretti}, and Mueller}]{todisco2015Design}
\bibinfo{author}{L.~Todisco}, \bibinfo{author}{C.~Fivet}, \bibinfo{author}{H.~{Corres-Peiretti}}, \bibinfo{author}{C.~Mueller},
\newblock \bibinfo{title}{Design and {{Exploration}} of {{Externally Post-Tensioned Structures Using Graphic Statics}}},
\newblock \bibinfo{journal}{JOURNAL OF THE INTERNATIONAL ASSOCIATION FOR SHELL AND SPATIAL STRUCTURES} \bibinfo{volume}{56} (\bibinfo{year}{2015}).
\bibitem[{{P{\'e}rez-Sala} et~al.(2018){P{\'e}rez-Sala}, Todisco, and {Corres-Peiretti}}]{perez-sala2018Exploration}
\bibinfo{author}{J.~C. {P{\'e}rez-Sala}}, \bibinfo{author}{L.~Todisco}, \bibinfo{author}{H.~{Corres-Peiretti}},
\newblock \bibinfo{title}{Exploration of externally post-tensioned spatial structures using {{3D}} graphic statics}  (\bibinfo{year}{2018}).
\bibitem[{Liu and Nocedal(1989)}]{liu1989Limited}
\bibinfo{author}{D.~C. Liu}, \bibinfo{author}{J.~Nocedal},
\newblock \bibinfo{title}{On the limited memory {{BFGS}} method for large scale optimization},
\newblock \bibinfo{journal}{Mathematical Programming} \bibinfo{volume}{45} (\bibinfo{year}{1989}) \bibinfo{pages}{503--528}. \DOIprefix\doi{10.1007/BF01589116}.
\bibitem[{Hammond et~al.(2011)Hammond, Jones, Lowrie, and Tse}]{hammond2011Embodied}
\bibinfo{author}{G.~Hammond}, \bibinfo{author}{C.~Jones}, \bibinfo{author}{E.~F. Lowrie}, \bibinfo{author}{P.~Tse},
\newblock \bibinfo{title}{Embodied {{Carbon}} - {{The Inventory}} of {{Carbon}} and {{Energy}} ({{ICE}})}  (\bibinfo{year}{2011}).
\bibitem[{Ching and Carstensen(2022)}]{ching2022Truss}
\bibinfo{author}{E.~Ching}, \bibinfo{author}{J.~V. Carstensen},
\newblock \bibinfo{title}{Truss topology optimization of timber--steel structures for reduced embodied carbon design},
\newblock \bibinfo{journal}{Engineering Structures} \bibinfo{volume}{252} (\bibinfo{year}{2022}) \bibinfo{pages}{113540}. \DOIprefix\doi{10.1016/j.engstruct.2021.113540}.
\bibitem[{Stern et~al.(2018)Stern, Wolf, and Mueller}]{stern2018Minimizing}
\bibinfo{author}{B.~Stern}, \bibinfo{author}{C.~D. Wolf}, \bibinfo{author}{C.~Mueller},
\newblock \bibinfo{title}{Minimizing {{Embodied Carbon}} in {{Multi-Material Structural Optimization}} of {{Planar Trusses}}},
\newblock in: \bibinfo{booktitle}{Proceedings of {{IASS Annual Symposia}}}, \bibinfo{publisher}{{International Association for Shell and Spatial Structures (IASS)}}, \bibinfo{year}{2018}.

\end{thebibliography}

\end{document}